\documentclass[aps,prb,floatfix,twocolumn,superscriptaddress,showpacs,10pt,longbibliography]{revtex4-1}
\usepackage{graphicx}
\usepackage{epstopdf}
\usepackage{textcomp} 
\usepackage{amsmath}
\usepackage{pstricks}
\usepackage{color}
\usepackage{hyperref}

\newcommand{\simlt}
      {\ifmmode       { \raisebox{-.8em}{$<$}\atop\sim}
         \else        {$\raisebox{-.8em}{$<$}\atop\sim$}
      \fi}
\newcommand{\mbf}{\mathbf} 
\renewcommand{\k}{{\mbf k}}
\newcommand{\q}{{\mbf q}}

\begin{document}
\title{Towards a quantitative description of tunneling conductance of superconductors: application to {LiFeAs}}
\author{A. Kreisel}
\affiliation{Niels Bohr Institute, University of Copenhagen, Universitetsparken 5, DK-2100 Copenhagen,
Denmark}
\affiliation{Institut f\" ur Theoretische Physik, Universit\" at Leipzig, D-04103 Leipzig, Germany}
\author{R. Nelson}
\affiliation{Institute of Inorganic Chemistry, RWTH Aachen University, Landoltweg 1, 52056 Aachen, Germany}
\author{T. Berlijn}
\affiliation{Center for Nanophase Materials Sciences, Oak Ridge National Laboratory, Oak Ridge, Tennessee 37831, USA}
\affiliation{Computer Science and Mathematics Division, Oak Ridge National Laboratory, Oak Ridge, Tennessee 37831, USA}
\author{W. Ku}
\affiliation{T. D. Lee Institute, Shanghai Jiao Tong University, Minhang, Shanghai 200240, China}
\affiliation{Department of Physics and Astronomy, Shanghai Jiao Tong University, Shanghai 200240, China}
\author{Ramakrishna Aluru}
\affiliation{Max-Planck-Institut f\"ur Festk\"orperforschung, Heisenbergstr. 1, D-70569 Stuttgart, Germany}
\affiliation{SUPA, School of Physics and Astronomy, University of St. Andrews, North Haugh, St. Andrews, Fife, KY16 9SS, United Kingdom}
\author{Shun Chi}
\affiliation{Department of Physics and Astronomy, University of British Columbia, Vancouver BC, Canada V6T 1Z1}
\affiliation{Quantum Matter Institute, University of British Columbia, Vancouver BC, Canada V6T 1Z4}
\author{Haibiao Zhou}
\affiliation{SUPA, School of Physics and Astronomy, University of St. Andrews, North Haugh, St. Andrews, Fife, KY16 9SS, United Kingdom}
\author{Udai Raj Singh}
\affiliation{Max-Planck-Institut f\"ur Festk\"orperforschung, Heisenbergstr. 1, D-70569 Stuttgart, Germany}
\author{Peter Wahl}
\affiliation{SUPA, School of Physics and Astronomy, University of St. Andrews, North Haugh, St. Andrews, Fife, KY16 9SS, United Kingdom}
\affiliation{Max-Planck-Institut f\"ur Festk\"orperforschung, Heisenbergstr. 1, D-70569 Stuttgart, Germany}
\author{Ruixing Liang}
\affiliation{Department of Physics and Astronomy, University of British Columbia, Vancouver BC, Canada V6T 1Z1}
\affiliation{Quantum Matter Institute, University of British Columbia, Vancouver BC, Canada V6T 1Z4}
\author{Walter N. Hardy}
\affiliation{Department of Physics and Astronomy, University of British Columbia, Vancouver BC, Canada V6T 1Z1}
\affiliation{Quantum Matter Institute, University of British Columbia, Vancouver BC, Canada V6T 1Z4}
\author{D. A. Bonn}
\affiliation{Department of Physics and Astronomy, University of British Columbia, Vancouver BC, Canada V6T 1Z1}
\affiliation{Quantum Matter Institute, University of British Columbia, Vancouver BC, Canada V6T 1Z4}
\author{P. J. Hirschfeld}
\affiliation{Dept. of Physics, U. Florida, Gainesville, FL 32611 USA}
\author{Brian M. Andersen}
\affiliation{Niels Bohr Institute, University of Copenhagen, Universitetsparken 5, DK-2100 Copenhagen,
Denmark}

\date{\today}

\begin{abstract}
Since the discovery of iron-based superconductors, a number of theories have been put forward to explain the qualitative origin of pairing, but there have been few attempts to make quantitative, material-specific comparisons to experimental results.  The spin-fluctuation theory of electronic pairing, based on first-principles electronic structure calculations,  makes predictions for the superconducting gap.  Within the same framework, the surface wave functions may also be calculated, allowing, e.g., for  detailed comparisons between theoretical results and measured scanning tunneling topographs and spectra.  Here we present such a comparison between theory and experiment  on the Fe-based superconductor LiFeAs.  Results for the homogeneous surface as well as impurity states are presented as a benchmark test of the theory.
For the homogeneous system, we argue that the maxima of topographic image intensity may be located at positions above either the As or Li atoms, depending on tip height and the setpoint current of the measurement.
We further report the experimental observation of transitions between As and Li-registered lattices as functions of both tip height and setpoint bias, in agreement with this prediction.
Next, we give a detailed comparison between the simulated scanning tunneling microscopy images of transition-metal defects with experiment.  Finally, we discuss possible extensions of the current  framework to obtain a theory with true predictive power for scanning tunneling microscopy in Fe-based systems.  
\end{abstract}


\maketitle

\section{Introduction}
The  Fe-based superconductor LiFeAs, with $T_c$ of $18\,\text{K}$, has lent itself particularly well to spectroscopic characterization by angular resolved photoemission spectroscopy (ARPES) and scanning tunneling microscopy (STM) due to the non-polar surface and the high quality of the samples. This enables a detailed comparison between spectroscopy and predictions by theory, and assessment of the status of the theoretical understanding of superconductivity in LiFeAs and iron-based superconductors in general. The gap structure of LiFeAs has been studied both by ARPES\cite{umezawa_unconventional_2012,Borisenko_12} as well as quasi-particle interference\cite{allan_anisotropic_2012}, which show a fourfold anisotropy of the gap and a characteristic distribution of gap magnitudes on different Fermi-surface sheets - yet neither of these yields information on the sign of the gap. There are a number of theoretical attempts to explain these results and in particular to calculate the detailed gap function, all of which led to the identification of sign-changing $s$-wave gaps\cite{Platt11,Wang13,Ahnetal14,Yin14,KontaniLiFeAs14,Uranga16}, but differed on the sets of Fermi-surface pockets that manifested the same sign. In part, this may be due to details of the low-energy band structure of LiFeAs, where there are several hole bands very close to the Fermi level, and electronic correlations are known to be important\cite{Ferberetal12,Haule12}.

A realistic description of the tunneling conductance as measured in STM has become attainable with the development of methods which account for the wave function overlap between the states of the tip and the surface\cite{Markiewicz09,Choubey14,Kreisel15,Demler16}. In this work, we present a detailed comparison between theoretical predictions for high-resolution spectra, topographs, and conductance maps with experiment. The results shed light on the current status of the progress towards a quantitative description of superconductivity in iron-based materials. 

The plan of this paper is as follows.  In Sec. II, we give details of the LiFeAs samples as well as of the measurement techniques.  In Sec. III, we present the theoretical framework used to analyze the data. In Sec. IV, we present our results for the homogeneous system and point out in particular that the interference of surface wave functions detected by the tip can lead to a change in registration of the conductance maxima at the Li or As lattices depending on tip height, setpoint bias, and other factors.  In Sec. V, we compare experimental results on pristine surfaces to these predictions.  In Sec. VI, we present calculations for impurity states and compare with experimental results.   In Sec. VII we  assess the success of the current framework and point out ways in which it could be improved.    Finally, in Sec. VIII we present our conclusions.

Some highlights of this work related to the impurity states were presented earlier in Ref. \onlinecite{Chi2016}.

\section{Experimental Details}
Experiments were performed in a home-built low temperature STM operating at temperatures down to $1.5~\mathrm{K}$ and in magnetic fields up to $14\mathrm T$ in cryogenic vacuum\cite{white_stiff_2011}. Samples were prepared by {\it in-situ} cleaving at low temperatures in cryogenic vacuum, resulting in atomically clean surfaces. We used tips cut from a PtIr wire. Bias voltages are applied to the sample, with the tip at virtual ground. Differential conductance spectra have been recorded through a lock-in amplifier with $f=413\mathrm{Hz}$ and a modulation of $V_\mathrm{mod}=500\mathrm{\mu V}$, unless stated otherwise. Data obtained in the superconducting state have been recorded at a temperature of $1.5\mathrm K$. Single crystals were grown using a self-flux method \cite{Chi_PRL}. Three transition-metal elements were substituted for iron separately, with the substitution levels of Mn $\sim 0.2\%$, Co $\sim 0.5\%$, and Ni $\sim 0.3\%$, respectively. Superconducting transitions determined by SQUID magnetometry show little changes with such minimal substitution levels comparing to non-substituted LiFeAs with $T_\mathrm c^{\mathrm{onset}} = 17$~K and a sharp superconducting transition $\Delta T \sim 1 $~K.

\section{Theory of homogeneous surface}
First we will establish a theoretical framework to calculate how the clean surface of the material is imaged by STM.

\subsection{Density functional theory calculations}

\begin{figure}[tb]
\includegraphics[width=1\columnwidth]{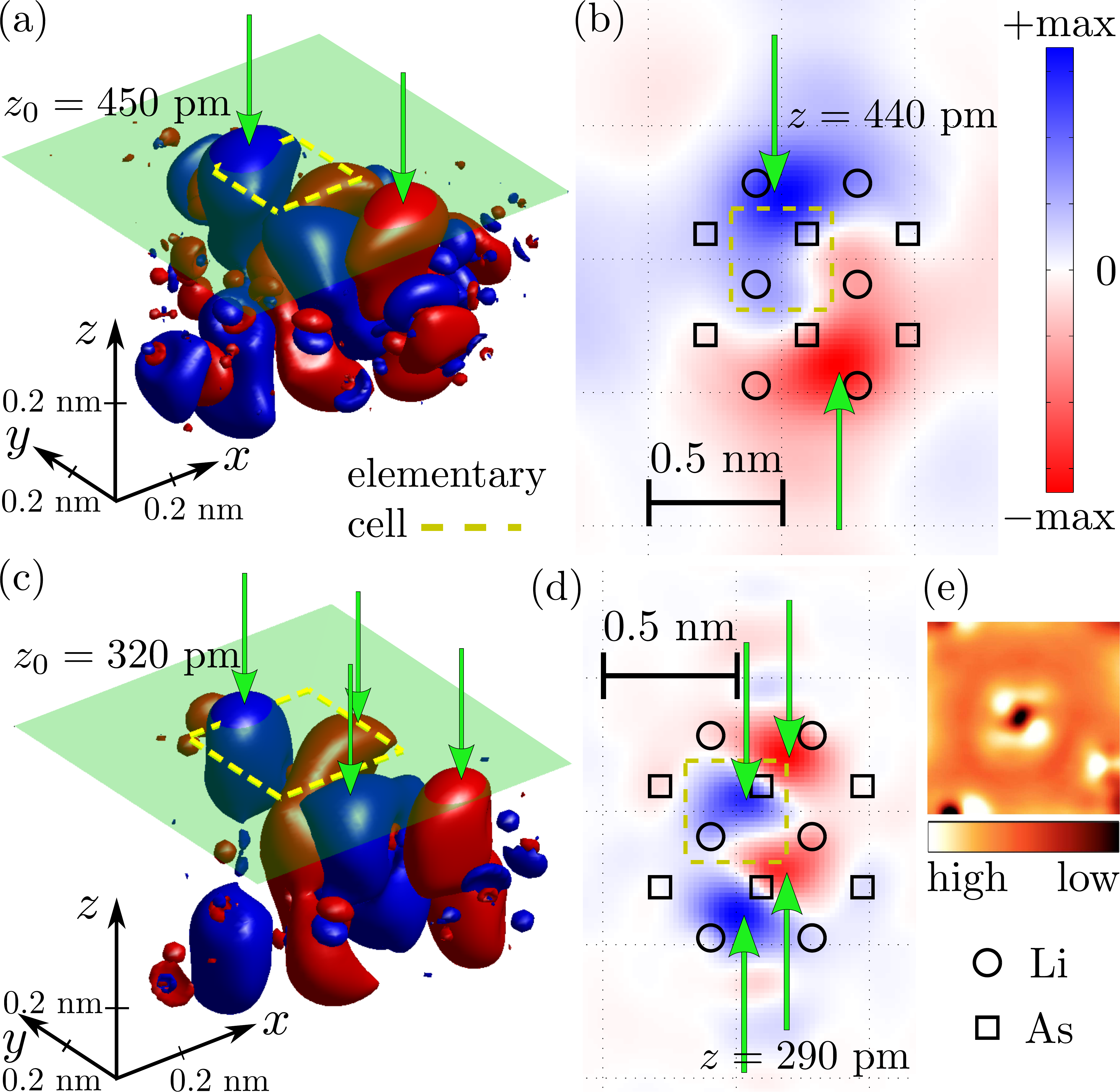}
\caption{Isosurface plots of a Fe-$d_{xz}$ Wannier orbital in LiFeAs at two different values (a, c) and corresponding plots of the same Wannier function at two different heights (b,d). The green plane at height $z_0$ in (a,c) is inserted for visual clarity of the isosurfaces.
 Red and blue indicate the phase of the wave function, see scale. The arrows point to the maxima of the Wannier function that gives rise to the maxima in the conductance maps whenever the weight of the  Fe-$d_{xz}$ Wannier function is large. Note that
the maxima move from positions above the atomic As positions (open squares) at small heights (e.g., large isovalues) to positions close to the atomic Li positions (open circles) at larger heights (e.g., smaller isovalues of the wave function). Experimental STM topography showing a native defect exhibiting chiral nature (e).}
\label{fig_wannier}
\end{figure}

The starting point is a first-principles calculation to obtain the material specific electronic structure.
In the present investigation we obtain the band structure 
using the Wien2K\cite{Blaha} package, followed by
projecting onto a  Wannier basis  preserving all local symmetries\cite{Ku_Wannier,Anisimov2005}; for details see Appendix \ref{appendix_dft}. The resulting five orbital tight-binding Hamiltonian of the electrons $c_{\mathbf R \sigma}^\dagger$ on the lattice is then given by
\begin{equation}
 H_{\text{TB}}=\sum_{\mathbf{R},\mathbf{R'}\sigma} t^{\mu\nu}_{\mathbf{R},\mathbf{R'}} c_{\mathbf R \mu  \sigma}^\dagger c_{\mathbf R' \nu \sigma},
 \label{eq_tb}
\end{equation}
where $t_{\mathbf{R},\mathbf{R'}}^{\mu\nu}$ are hopping elements between orbitals $\mu$ and $\nu$ on Fe sites labeled with $\mathbf{R}$ and $\mathbf{R'}$.
For convenience and later reference, we denote matrix quantities of size of the number of orbitals, e.g., $5\times 5$ matrices, with a hat. Thus, for example, the normal-state Hamiltonian matrix in momentum space reads
\begin{equation}
 \hat H(\mathbf{k})^{\mu \nu}=\sum_{\mathbf{\delta}}e^{i\mathbf{k}\cdot\mathbf{\delta}}t_{\delta}^{\mu\nu},
 \label{eq_H_normal}
\end{equation}
where $\delta={\mathbf{R}-\mathbf{R'}}$ is the real-space distance of the hopping $t_{\mathbf{R},\mathbf{R'}}^{\mu\nu}$.
Since the electronic structure of LiFeAs (and of Fe-based superconductors in general) exhibits only limited dispersion in the $k_z$ direction, we consider a model at $k_z=0$ in the following. 
The {\it ab initio} approach also gives us a set of Wannier orbitals, one of them shown in Fig.~\ref{fig_wannier}.
Note that the symmetry of the Wannier functions is only constrained by the space-group symmetry of the crystal structure, e.g., the overall symmetry is lower than for $d$ states in a cubic environment.
Furthermore, individual Wannier functions typically exhibit significant weight on nearby atoms as well.

It is interesting to note from Fig.~\ref{fig_wannier}(b,d) the chiral nature of the $d_{xz}$ orbital shown, since certain native defects in LiFeAs are known to exhibit such a chiral topographic structure. We do not study these defects here, but note that local orbital ordering that distinguishes between $d_{xz}$ and $d_{yz}$\cite{Inoue12} could lead to a chiral defect map (for comparison see Fig.~\ref{fig_wannier}(e) for an STM image of such a defect).

\begin{figure}[tb]
 \includegraphics[width=0.99\linewidth]{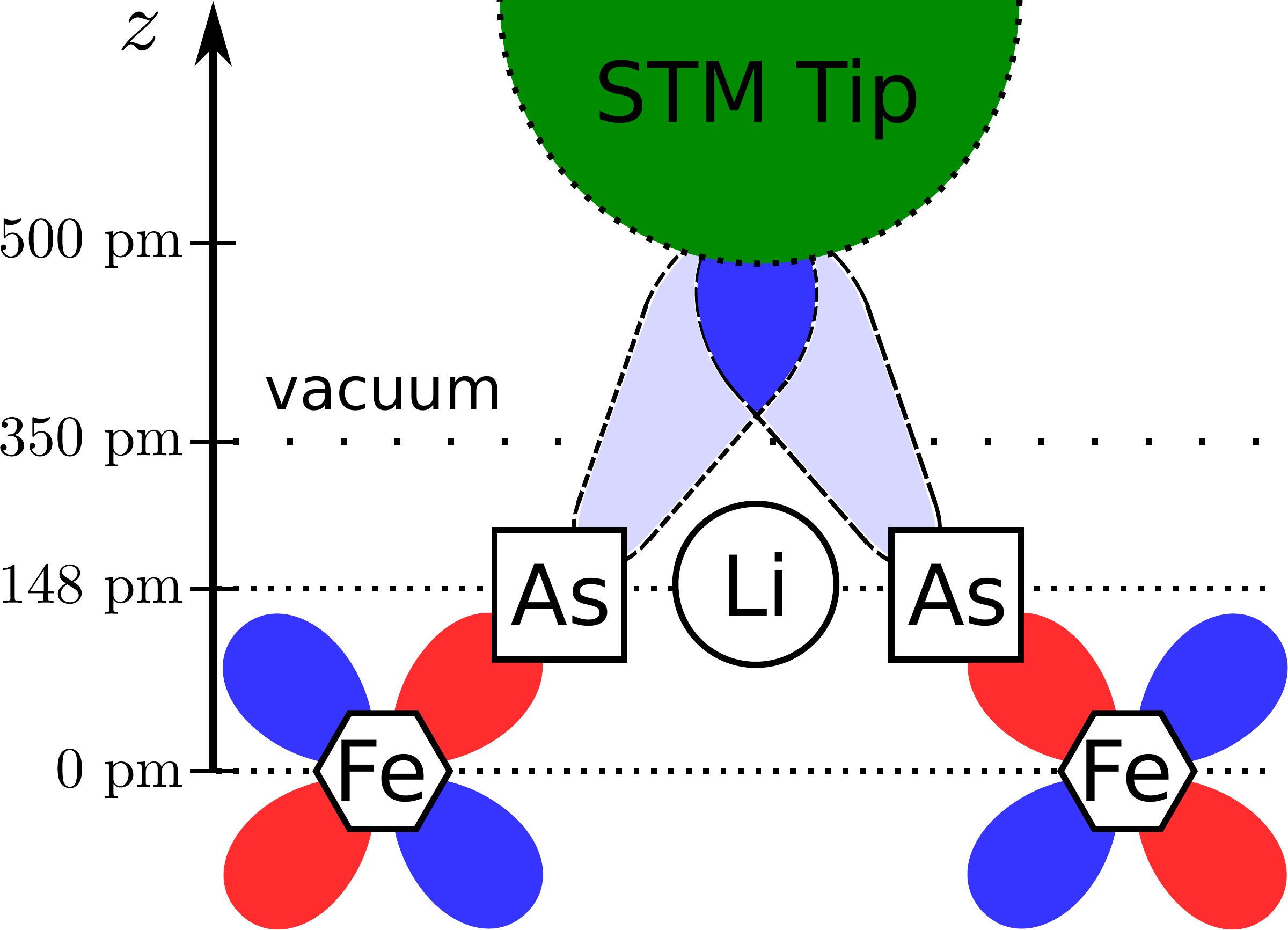}
\caption{Schematic illustration of the tunneling process indicating how tunneling into As states can lead to an intensity maximum above the Li positions: Interference between electrons in Wannier states of neighboring Fe atoms (phase of wave function symbolized by red/blue color) can give rise to maximal tunneling when the STM tip is located above the Li atom. A tip closer to the surface will shift the maximal tunneling positions towards the As atoms.}
\label{fig_cartoon}
\end{figure}

The As $p$ states  form the major contribution to the Fe-3$d$ Wannier function shown at several \AA~ above the surface of the sample, where the tip is positioned. As seen in Fig.~\ref{fig_wannier}, they have a rather large extent, such that the maximum values of the function {\it sufficiently far above the surface} occur closer to the Li sites rather than the As sites, as shown schematically in Fig. \ref{fig_cartoon}. It is important to recognize that since the Li states are quite far from the Fermi level, they have negligible contributions in the Fe-3$d$ bands, rendering the Li itself effectively invisible for STM.  Nevertheless there are consequences for STM, which is sensitive to the wave functions several \AA~ above the surface, leading to possible misidentification of atomic lattice positions, as discussed further below.  

\subsection{Superconducting gap}

The superconducting gap of the homogeneous system is obtained by the self-consistent solution of the Bogoliubov-de Gennes (BdG) equation in momentum space.
For this purpose, the Nambu Hamiltonian
\begin{equation}
 \underline{\hat H}=\left(\begin{array}{cc}
              \hat H(\mathbf{k})&\hat \Delta(\mathbf{k})\\
              \hat\Delta(\mathbf{k})^T&-\hat H(\mathbf{-k})
                  \end{array}\right),
                  \label{eq_H_nambu}
\end{equation}
is diagonalized on a $k$ grid (of typical size $32\times32$), yielding the eigenvalues $\{\pm E_l(\mathbf{k})\}$ and the unitary transformation
$\underline{\hat U}(\mathbf{k})=(\hat u_{\mathbf{k}},\hat v_{\mathbf{k}};\hat v_{\mathbf{k}},\hat u_{\mathbf{k}})$ that diagonalizes $\underline{\hat H}$, a $10\times 10$ matrix indicated by the underscore.
The  gap in orbital space is calculated from the self-consistency relation
\begin{equation}
 \hat \Delta(\k)^{\mu\nu}=\sum_{\mu^\prime,\nu^\prime}\sum_{\mathbf{k^\prime},l}\Gamma_{\mu^\prime\mu\nu\nu^\prime}(\k,\k')u_{\mu^\prime}^lv_{\nu^\prime}^l f(E_l(\mathbf{k})),
 \label{eq_gap_scf}
\end{equation}
where $u_{\mu^\prime}^l$ ($v_{\nu^\prime}^l$) are the elements of the column vectors 
$\hat u_{\mathbf{k}}$ ($\hat v_{\mathbf{k}}$) and $f(E)$ denotes the Fermi function.
The pairing interactions $\Gamma_{\mu^\prime\mu\nu\nu^\prime}(\k,\k')$ are derived from standard spin-fluctuation theory\cite{a_kemper_10} together with a proper symmetrization in the spin-singlet channel\cite{Roemer15}  as outlined in Appendix \ref{appendix_spin_fluc}.
For a visualization of the superconducting gap in band space, the normal-state transformation that diagonalizes $\hat H(\mathbf{k})$ of Eq.~(\ref{eq_H_normal}) from orbital space to band space is used, yielding the gap structure shown in Fig.~\ref{fig_delta} of Appendix \ref{appendix_spin_fluc}.

\begin{figure}[tb]
\includegraphics[width=\linewidth]{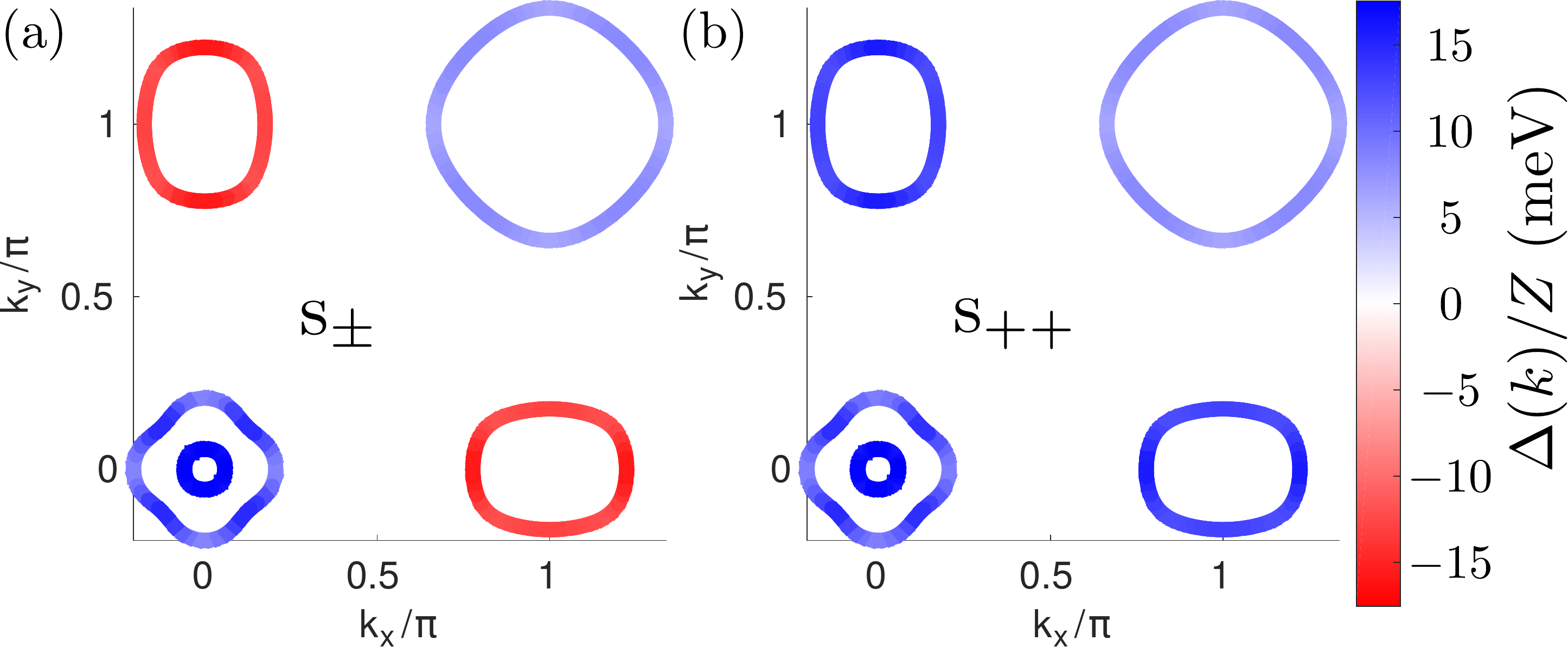}
\caption{(a) Superconducting gap from the self-consistent BdG equation, plotted at the positions of the Fermi surface in momentum space. (b) For the $s_{++}$ gap function we use the same gap structure (thus giving rise to the same DOS), but without the sign change.}
 \label{fig_gap}
\end{figure}

In order to provide a clear picture, the gap structure is plotted on the $k_z=0$ Fermi surface in Fig.~\ref{fig_gap}(a) where the sign changing $s_\pm$ symmetry is explicitly seen. In agreement with conclusions drawn from a variety of experiments, the gap function is nonzero everywhere on the Fermi surface, and the gap on the outer, $\Gamma$-centered $d_{xy}$ pocket appears to be the smallest in magnitude.  Because the density functional theory (DFT) Fermi surface is not consistent with ARPES experiments, which find tiny or nonexistent inner $d_{xz,yz}$ hole pockets, the gap function cannot be regarded as a completely correct representation of the true gap in the system.  Nevertheless the relative magnitudes of the gaps appear qualitatively quite reasonable compared with ARPES.\cite{Borisenko_12, umezawa_unconventional_2012}

\begin{figure}[tb]
 \includegraphics[width=\linewidth]{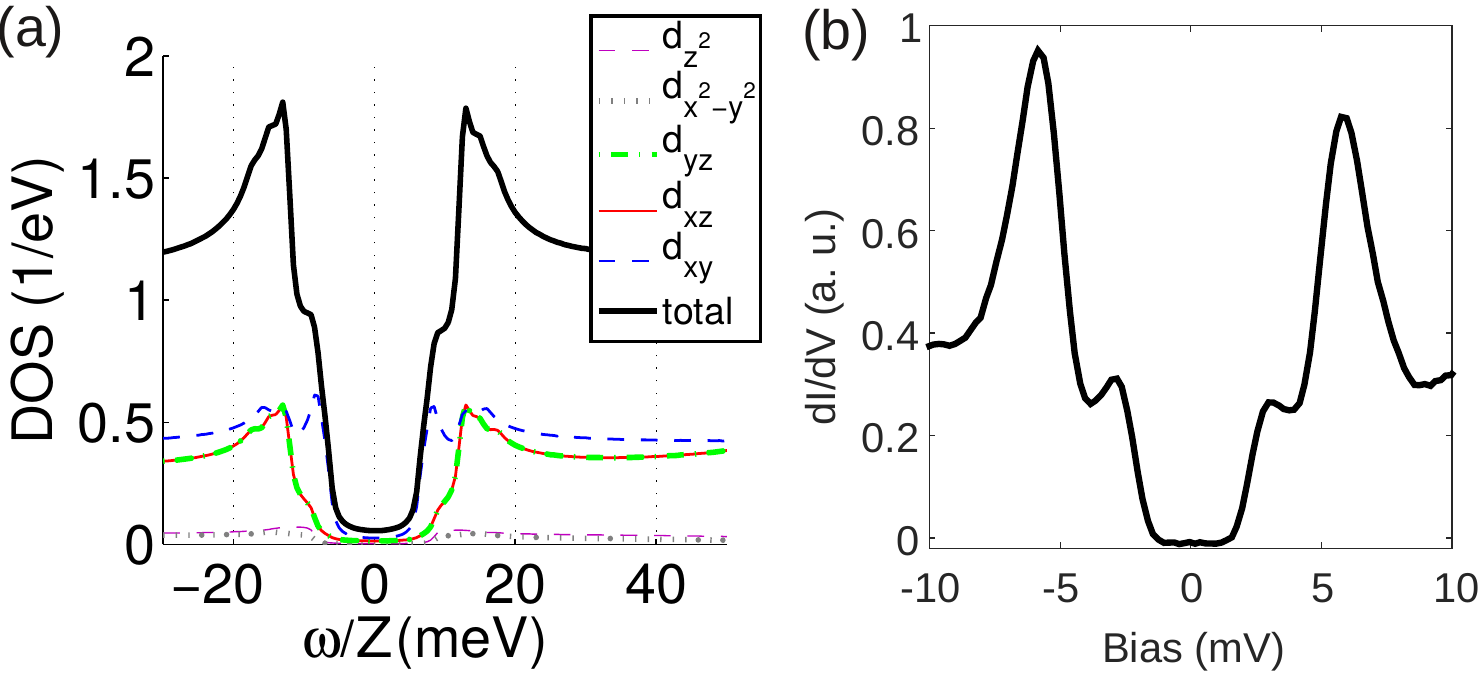}
\caption{(a) Orbital resolved DOS of the homogeneous superconductor. The orbital content changes strongly between the energy scales $10\,\text{meV}$ and $20\,\text{meV}$ due to the gap structure. Note renormalization of DFT energy scale by $Z$ as discussed in text. (b) Experimental STM spectrum showing the differential conductance of homogeneous LiFeAs. ($V_s=15\mathrm{mV}$, $I=150\mathrm{pA}$,T = 1.6K).}
 \label{fig_orbital_dos}
\end{figure}

For the calculation of the density of states (DOS) in the superconducting state we construct the matrix Green's function
 \begin{equation}
  \underline{\hat G}(\mathbf{k},\omega)=(\omega -\underline{\hat H}(\mathbf{k})+i\eta)^{-1},
 \end{equation}
from the self-consistent solution of the BdG equation,
which allows us to calculate the orbitally resolved homogeneous DOS $\rho^\mu(\omega)=-\frac 1\pi \mathop{\text{Im}}\sum_{\mathbf{k}}\hat G(\mathbf{k},\omega)^{\mu\mu}$, manifesting
a two gap feature at low energies and a prominent change of  weight from mainly
$d_{xy}$ within the large gap feature, to dominantly $d_{xz}$/$d_{yz}$ starting from
energies of the order of the coherence peak, see Fig. \ref{fig_orbital_dos}. By a Fourier transform,
we obtain the real-space {\it lattice} Green's function
\begin{equation}
 \underline{\hat G}_{\mathbf{R}, \mathbf{R'}}^0(\omega)=\sum_{\mathbf{k}}e^{-i\mathbf{k}\cdot(\mathbf{R}-\mathbf{R'})}
\underline{\hat G}(\mathbf{k},\omega)=\underline{\hat G}_{\mathbf{R}-\mathbf{R'}}^0(\omega)\,,\label{eq_lattice_GF}
\end{equation}
which we introduce for later reference.

Note that the gaps in our approach appear to give an LDOS spectrum for the homogeneous system which agrees well with the conductance spectrum measured in STM, as seen from Fig.~\ref{fig_orbital_dos}(b). Our results are generally quite similar to those found in Ref. \onlinecite{gastiasoro_impurity_2013}, and show a small gap feature arising from the $d_{xy}$ states and a larger gap derived from the $d_{xz,yz}$ states, with relative magnitude consistent with experiment. Note that the gap structure as calculated from the full BdG equation is also very similar to the result of the solution of the linearized gap equation with the present two dimensional band structure and a result obtained from a three dimensional DFT derived band structure when plotted in the $k_z=0$ plane\cite{Wang13}. Finally, we note that correlations will give rise to a renormalization factor $Z$ which essentially changes the overall energy scale in the calculation. Setting $Z=1/2$ then matches the experimental magnitude of the superconducting gap and roughly agrees with observed renormalizations of the electronic structure in the normal state\cite{Ferberetal12,Haule12,Borisenko_12}

\subsection{Wannier functions to calculate tunneling current}

The differential tunneling conductance in an STM experiment at a given bias voltage $V$ is given by\cite{Tersoff1985}
\begin{equation}
\frac{dI}{dV}(\mathbf{r},eV)=\frac{4\pi e}{\hbar} \rho_t(0) |M|^2 \rho(\mathbf{r},eV),
\label{eq_conductance}
\end{equation}
where $\mathbf{r}=(x,y,z)$ denote the coordinates of the tip, $\rho(\mathbf{r},\omega)$ is the continuum LDOS (cLDOS),
$\rho_t(0)$ is the DOS of the tip, and $|M|^2$ is the square of the matrix element for the tunneling barrier.
The cLDOS can be calculated by
\begin{equation}
 \rho(\mathbf{r},\omega)\equiv -\frac 1 \pi\mathop\text{Im}\underline{G}^{11}(\mathbf{r},\mathbf{r};\omega ),
 \label{eq_cLDOS}
\end{equation}
where $\underline{G}^{11}(\mathbf{r},\mathbf{r};\omega )=G_\uparrow (\mathbf{r},\mathbf{r};\omega )$ is the normal part of the Nambu continuum Green's function given by
\begin{equation}
\underline{G}(\mathbf{r},\mathbf{r}';\omega )=\left(
\begin{array}{cc}
G_\uparrow (\mathbf{r},\mathbf{r}';\omega ) & F(\mathbf{r},\mathbf{r}';\omega )\\
F^*(\mathbf{r},\mathbf{r}';\omega )&-G_\downarrow (\mathbf{r},\mathbf{r}';-\omega )
                                             \end{array}
                                             \right),
                                             \label{eq_nambu_def}
\end{equation}
defined as usual using the field operators $\psi_\sigma(\mathbf{r})$.
These are related to the lattice operators defined earlier via
\begin{equation}
\psi_\sigma(\mathbf{r})=\sum_{\mathbf{R} \mu } c_{\mathbf{R}\mu\sigma}w_{\mathbf{R}\mu}(\mathbf{r}),
\end{equation}
where the first-principles-derived Wannier functions $w_{\mathbf{R}\mu}(\mathbf{r})$ are the matrix elements.

Employing the Wannier basis transformation, we can compute the continuum Green's function\cite{Choubey14} by the expression
\begin{equation}
 \underline{G}(\mathbf{r},\mathbf{r}';\omega)=\sum_{\mathbf{R}, \mathbf{R'},\mu\nu}\underline{\hat G}_{\mathbf{R}, \mathbf{R'}}^{\mu,\nu}(\omega)w_{\mathbf{R}\mu}(\mathbf{r})w_{\mathbf{R'}\nu}(\mathbf{r'}),
 \label{eq_basis}
\end{equation}
and finally obtain the conductance from Eq.~(\ref{eq_conductance}).
Note that the cLDOS $\rho(\mathbf{r},\omega)$ involves
not only local lattice Green's function contributions $\underline{\hat G}_{\mathbf{R},\mathbf{R}}(\omega)$, but also nonlocal contributions with $\mathbf{R}\ne\mathbf{R'}$.
Due to this, the spectral and spatial properties of the cLDOS
are typically both quantitatively and qualitatively different from those of the lattice DOS.
We stress that Eq.~(\ref{eq_conductance}) for the conductance in a STM experiment only holds under the assumptions that the tip is sharp enough such that only one atom contributes to the tunneling process. Below we explore the variability of surface imaging within the scope of this ansatz.

\begin{figure}[tb]
 \includegraphics[width=\linewidth]{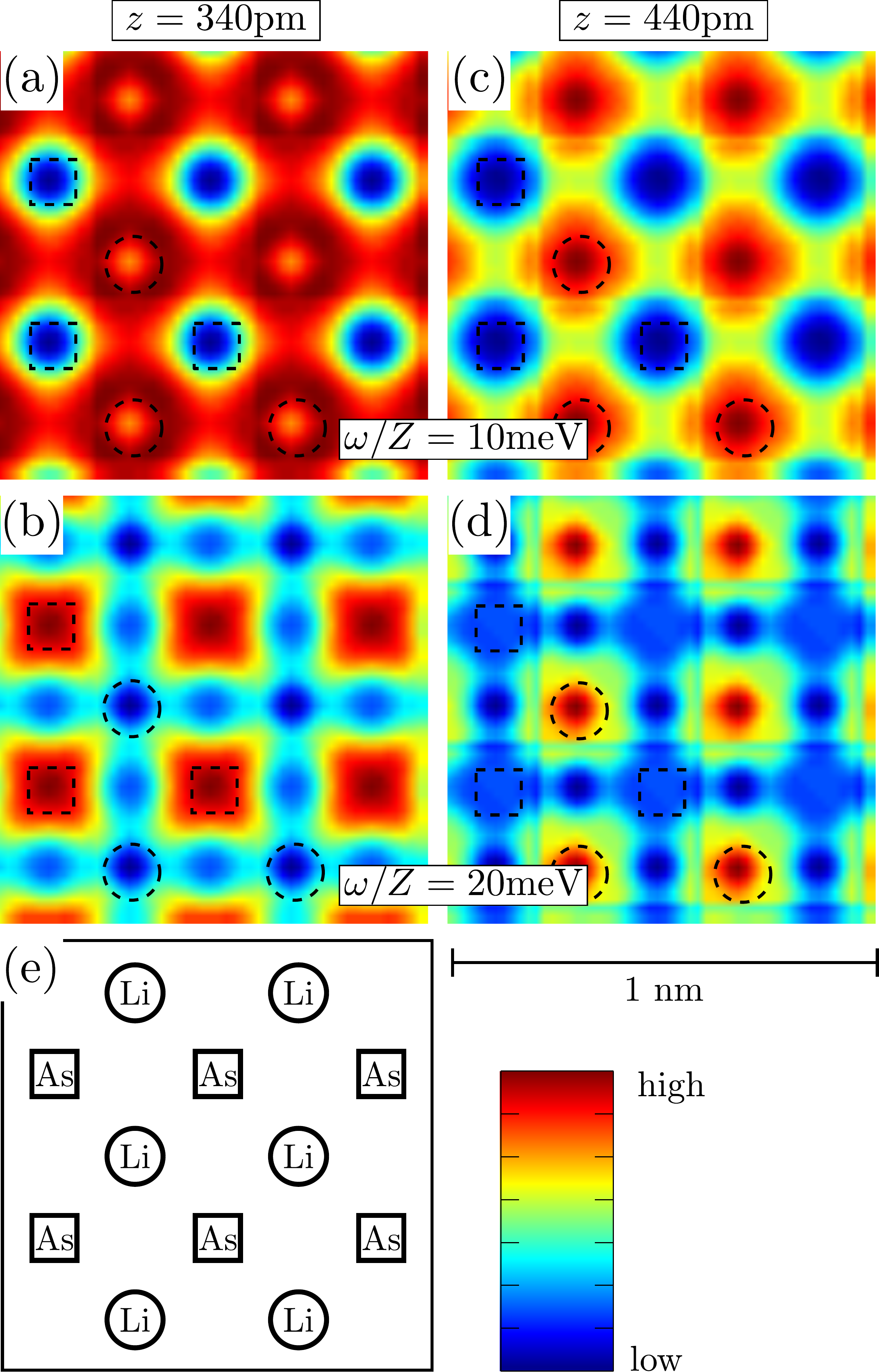}
\caption{Continuum LDOS calculated at different heights (left column, (a-b): 340 pm, right  column, (c-d): 440 pm) above the surface, energies as indicated on the plots ($\omega/Z=10\,\text{meV}$, $\omega/Z=20\,\text{meV}$). While for the small tip height
the positions of the maxima change from the surface As positions at high energies to the Li positions at small energies, there is no such change when the tip is at larger distances from the surface. All plots are with the same field of view covering an area of $1\times 1 \,\text{nm}^2$ with the atomic positions of the surface atoms as marked in (e).  Lines along the crystallographic axis in the calculated maps are artifacts from using a Wannier function that has a finite extension in the plane. These become stronger at larger heights because the Wannier function is less peaked, see Fig. \ref{fig_wannier}.}
\label{fig_hom}
\end{figure}

\section{Results: Homogeneous system}
In this section, we use the lattice Green's function, Eq.~(\ref{eq_lattice_GF}) of the homogeneous system in Eq.~(\ref{eq_basis}) and calculate the cLDOS using Eq.~(\ref{eq_cLDOS}).
In Fig.~\ref{fig_hom}, we show the corresponding differential conductance maps of the homogeneous system calculated at different tip heights.  Using the cLDOS, we can access intra-unit-cell structure  reflecting the properties of the electronic wave functions above the surface, in contrast to traditional calculations in Fe-based superconductors restricted to the lattice LDOS in the Fe plane (which is of course independent of the lattice position $\mathbf{R}$ in the homogeneous case).
As shown in Fig.~\ref{fig_hom}, changing the height of the STM tip may change the positions of the conductance  maxima, as already suggested  by Figs.~\ref{fig_wannier} and \ref{fig_cartoon}, and this change may be bias dependent. At large height, i.e.,  small tunneling current,
the maxima are above the positions of the Li atoms for all setpoint voltages shown. The reason is the constructive interference of the Fe Wannier functions from nearby Fe atoms, as shown schematically in Fig. \ref{fig_cartoon}. On the other hand, the small height cLDOS maps show maxima at positions of As atoms on the surface when the setpoint voltage is outside the superconducting gap, and maxima close to the Li positions are observed for energies below the large gap.

This switching can be understood in terms of the dominant orbital weight at these energies in conjunction with the form of the Wannier functions. At large energies ($\omega/Z > 15 \,\text{meV}$), the orbital weight is $d_{xz}/d_{yz}$ and $d_{xy}$, see Fig.~\ref{fig_orbital_dos}. In the maps at large tip height, one sees that the $d_{xz}$ Wannier function is actually maximal close to the next-nearest-neighbor (NNN) Li positions (see Fig. \ref{fig_wannier}), such that constructive interference will give enhanced contributions at these positions leading to the maxima as observed. At small energies,  the  $d_{xy}$ orbital dominates the DOS, but it is also larger
at the NNN Li positions than at the NNN As positions.

The same happens at lower tip height in the low-energy regime.
However with increasing energy, e.g., including also contributions of the $d_{xz}/d_{yz}$ Wannier functions, the maxima move towards the As positions since the $d_{xz}/d_{yz}$ Wannier functions {\it at the lower height} are larger at the nearest-neighbor (NN) As positions than at the NNN Li positions, see Fig. \ref{fig_wannier}. These results are robust in terms of the approximations made for the calculations since
the Wannier functions represent largely high-energy properties of the system, i.e., should be well described by the first-principles approach.

\begin{figure}[tb]
 \includegraphics[width=\linewidth]{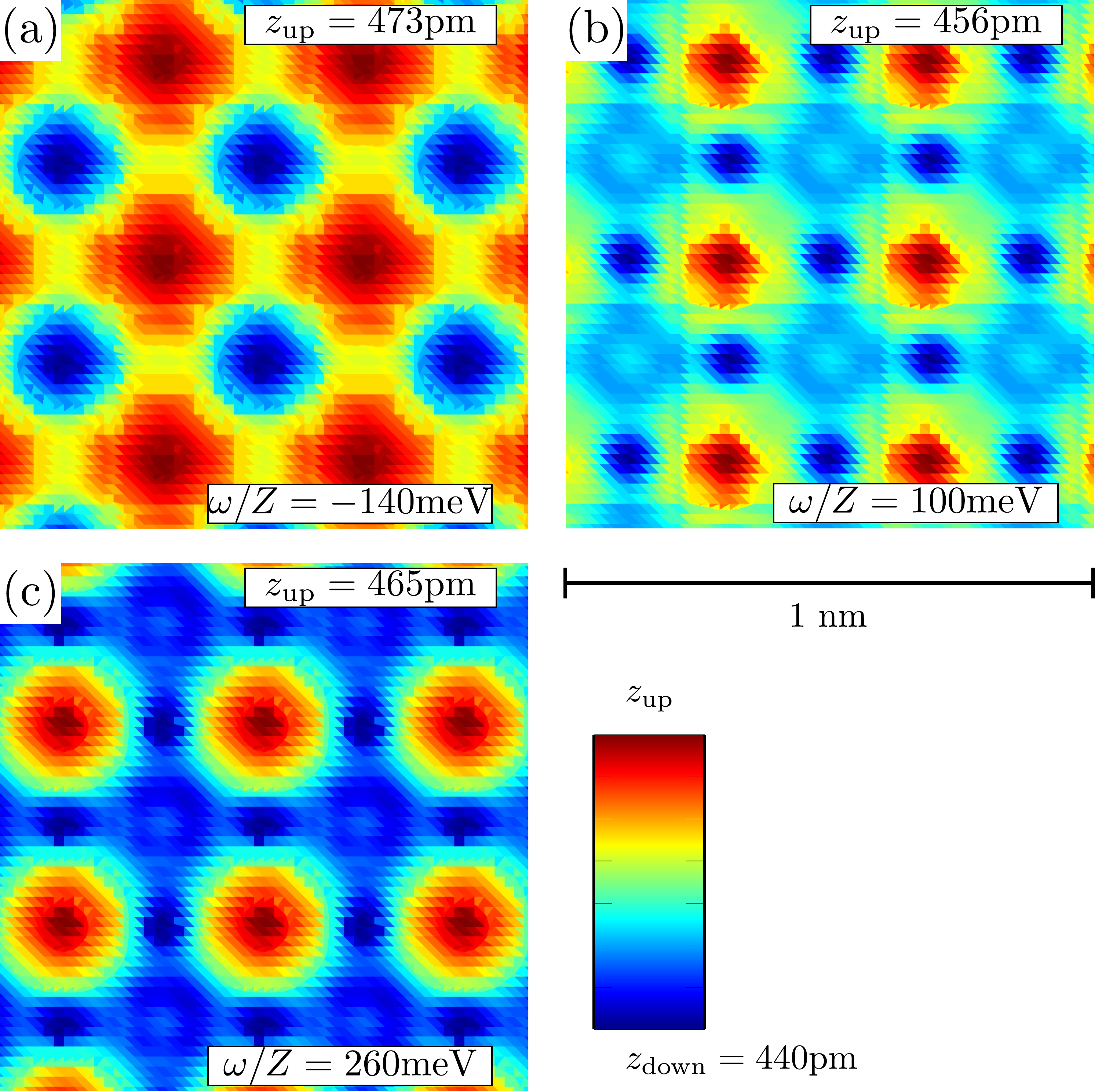}
\caption{Topographs of an area of $1\times 1\,\text{nm}^2$  calculated such that the closest point of the tip is fixed to $z_{\mathrm{down}}=440\,\text{pm}$ (measured from the Fe plane, see Fig. \ref{fig_cartoon}) for different bias voltages (a-c) as indicated.
Note that for huge positive setpoint bias voltages, the maximal heights are above the As atoms in the lattice (c), whereas for smaller voltages the maximal heights are registered above the Li atoms (a,b).
For even smaller setpoint bias voltages, e.g., within the large superconducting gap, the maxima are again found  above the As positions, see Fig. \ref{fig_hom} (e). Artifacts of the finite range
of the Wannier functions used in the calculations are present, mostly visible in (b). See Fig. \ref{fig_hom} for the relative lattice position of the surface atoms in the field of view. }
\label{fig_topography_huge}
\end{figure}

Whether the switching of the lattice can be observed experimentally via STM, may well depend on the properties of the STM tip which we assumed until now to be a small tip without structure\cite{Tersoff1985}.  A more realistic description of the tunneling process would involve convolving the surface wave functions with the wave function of the tip.  
While we do not discuss the effects of tip orbital structure here, it is clear that even assuming
a spherical symmetric tip wave function\footnote{Strictly speaking this assumption is only correct for a STM tip made of a s-wave metal.
Note, however, that also metals with $d$ electrons at the Fermi level can have a spherical symmetric overall tip wave function, if crystal fields are neglected.}
some smearing of the cLDOS will occur when calculating the actual conductance maps.
If this smearing due to the tip wave function is too large,  the switching
of the  lattice registration from Li to As  with bias seen in
Fig. \ref{fig_hom} will be removed, and the maxima will always be observed above the Li sites until the tip is unphysically close to the surface. Lattice registration switchings with bias of this sort discussed here have been seen occasionally in experiment, but are rarely reported. One exception is Ref. \onlinecite{SchlegelPhD} where the spatial positions of the maxima of the tunneling current are seen to switch as a function of bias voltage and contrast asymmetries with respect to the sign of the voltage are discussed. 

In order to simulate images of topographs, we calculate the cLDOS $\rho({\bf r},eV)$ at several heights and energies, then solve
\begin{equation}
I_0=\frac{4\pi e}{\hbar}\rho_t(0) |M|^2\int_0^{eV} d\omega~  \rho({x,y,z(x,y)},\omega)\,,\label{eq:topograph}
\end{equation} at setpoint current $I_0$ for the height map $z(x,y)$.

We obtain topographs that
show the same switching of the observed positions of the maxima (now in $z$), as shown in Fig.~\ref{fig_topography_huge}.
The effect is, however, less pronounced due to the integration over energy which adds up contributions
with maxima at Li positions and As positions, giving rise to a relative flat result.
Topographs for large bias away from the superconducting gap magnitude can however
 still give rise to switching of the positions as seen from Fig.~\ref{fig_topography_huge}.

\begin{figure*}[tb]
\includegraphics[width=\linewidth]{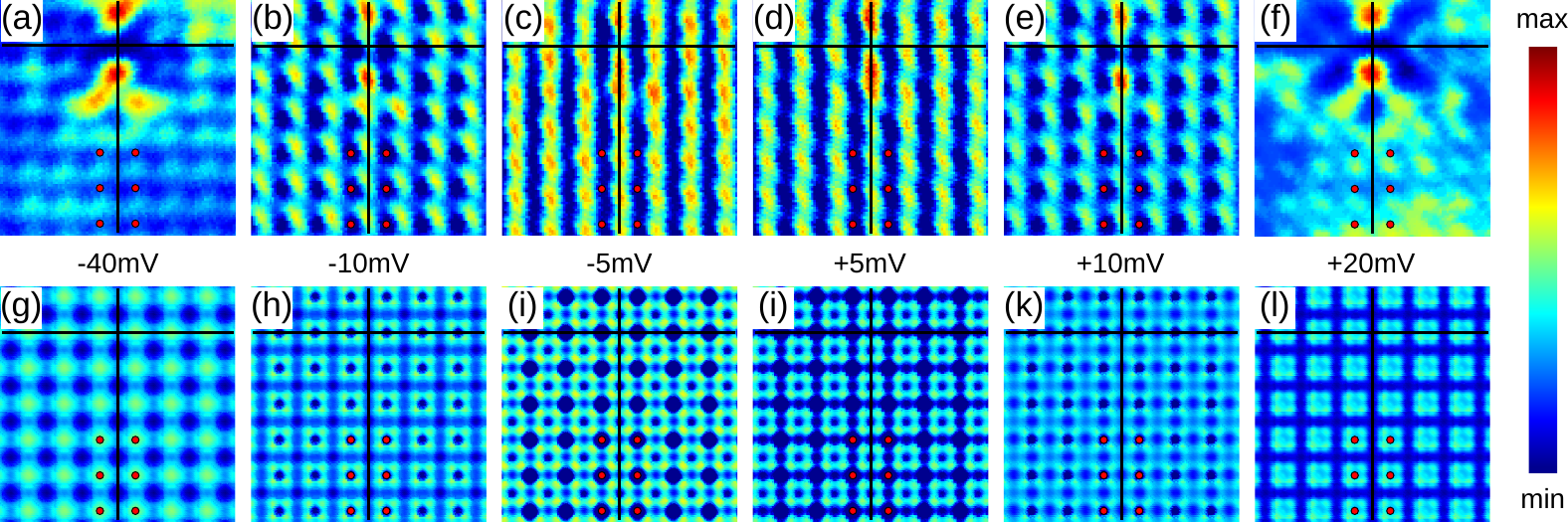}
\caption{(a-f) Topographic experimental STM images taken with different bias voltages with the same current $I=2\,\mathrm{nA}$. All images show the same defect and have been taken with the same tip configuration. The small red dots label the same lattice positions relative to the impurity; for the theoretical calculations (g-l), these are the positions of the Li atoms. The black lines cross at the center line of the defect state (Fe position), the position of which is fixed and can be used as a reference. A phase shift is observed in the atomic lattice when the bias voltage and tip-sample distance is reduced, compare e.g., panels (a,f) to panels (b-e). A similar shift is found in the theoretical simulations within the same energy range  where the cross also marks a Fe position. Note that for the theoretical calculations, the energy has been renormalized by a factor of $Z=1/2$ consistent with all other calculations in this and previous work \cite{Chi2016}.
}
\label{lattice_shift}
\end{figure*}

\begin{figure}[t]
\includegraphics[width=1\columnwidth]{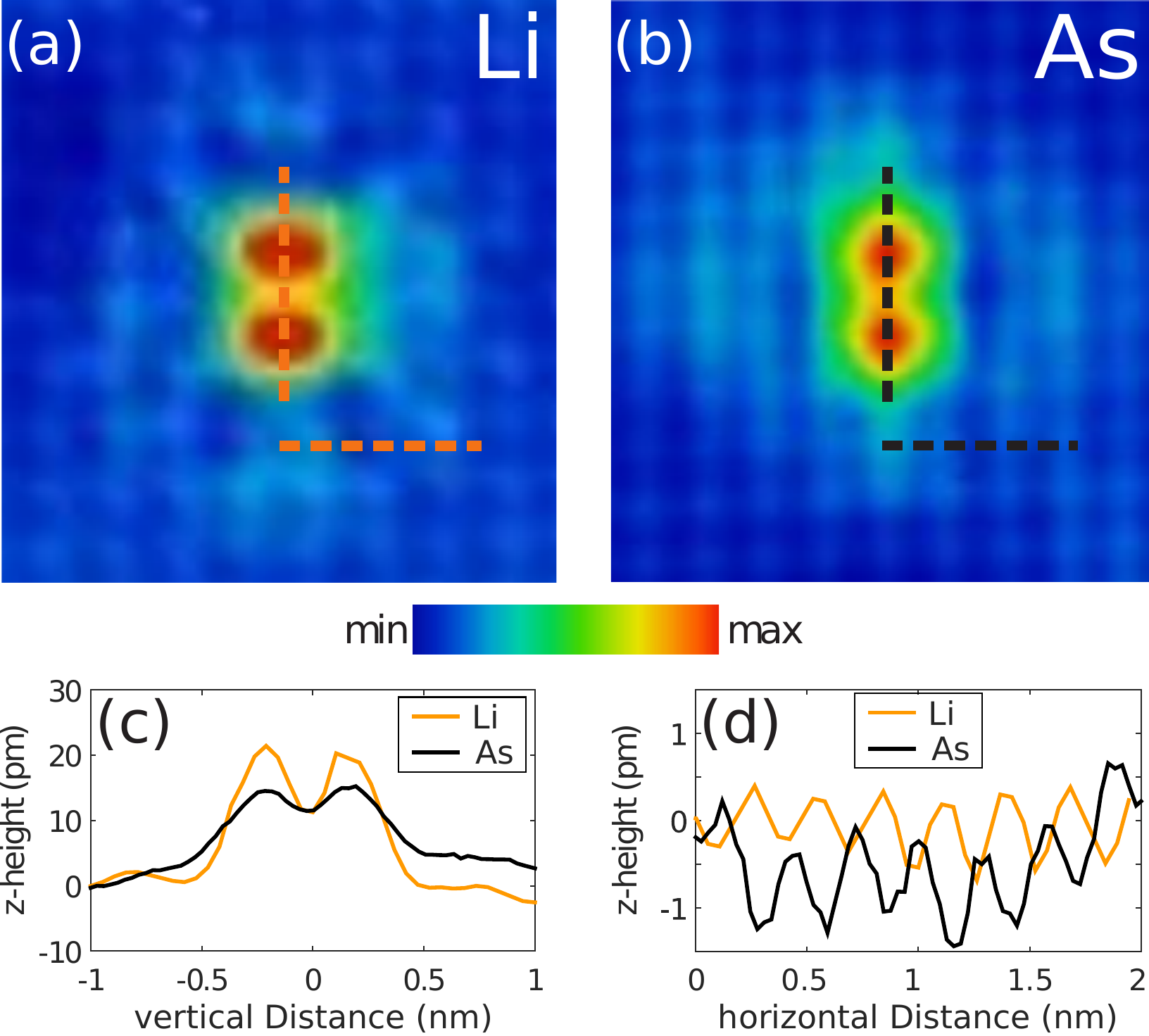}
\caption{Imaging the atomic lattice in STM with different tips;
 (a) STM topography showing a single Mn impurity with maxima imaged at the position of lithium (Li) atoms. ($V_s=-25\mathrm{mV}$, $I=100\mathrm{pA}$). The impurity here serves merely as a point of reference. (b) STM topography of the same Mn impurity obtained with a different apex of the tip, imaging maxima at the positions of arsenic (As) atoms. ($V_s=-50\mathrm{mV}$, $I=300\mathrm{pA}$). Topographs are imaged in the normal state at $B = 10\mathrm T$ and $T = 12\mathrm K$. (c) height profile of the Mn defect along the vertical dashed lines indicated in (a,b) and (d) height profile of the atomic modulation of the homogeneous system extracted along the horizontal lines in (a,b). The atomic resolution is found to be in anti-phase between the two topographs.}
\label{fig_exp_lattice}
\end{figure}

\section{Comparison with Experiment}
As stated above, there are no systematic studies of these effects reported in the literature. Here, we show a series of topographic STM images taken at different bias voltages clearly revealing a phase shift between images taken at large bias voltage and small bias voltage (see Fig. \ref{lattice_shift}), consistent with the theoretical prediction.
Note that the theoretical calculations have been done just for the homogeneous system Fig. \ref{lattice_shift} (g-l) and show maximal height at the positions of the Li atoms when the absolute bias voltage is larger than $30\,\text{mV}$, while the height maxima gradually move to As positions for smaller bias voltage. All simulations have been done for fixed current conditions such that the $z$-position gets reduced at small bias. This behavior can be understood in the picture of the interference as sketched in Fig. \ref{fig_cartoon}. Experimentally, the switching is not as clean since the topographs show some deviations from a non-spherical tip, but by help of the fixed position of  the same impurity, it can be seen that the maxima clearly move horizontally as a function of bias voltage in qualitative agreement to the simulations.

It is worth noting though that not all tip configurations exhibit this shift. To illustrate the effect on the Li/As lattice imaging due to tip characteristics,  we display in Fig. \ref{fig_exp_lattice} experimental topographs obtained with two different tips near a Mn defect which serves only as a point of reference. While both tips image the surface and the defect with high spatial resolution, despite the rather similar tunneling parameters, they image the surface atomic structure once with maxima on the Li atoms and once on the As lattice. From the theoretical analysis, this suggests that the tip orbital probes different electronic states of the surface, as is also evident from the scattering patterns detected in the images.

\section{Impurity states}
\subsection{Calculation of impurity potentials}
To obtain the impurity potentials of transition-metal substituents for Fe from first-principles, we make
use of the  Wannier function based effective Hamiltonian method for disordered systems\cite{Berlijn11,Berlijn12,Choubey14} The influence of the TM substitutions on the Hamiltonian is extracted from two DFT  calculations: the undoped LiFeAs and the impurity supercell system
Li$_8$Fe$_7$TMAs$_8$, according to the procedure described in Ref. \onlinecite{Berlijn11}.\cite{Choubey14}
The orbitally resolved potentials for Ni, Mn, and Co are summarized in Table \ref{table_Imp} of  Appendix B and are very similar in magnitude to those calculated earlier for a different compound\cite{nakamura_first-principles_2011}. The orbital dependence of the impurity potentials is not very pronounced, but has been taken into account for the following calculation of impurity states.

\subsection{T-matrix approach}

Including an impurity term in the Hamiltonian
\begin{equation}
 H_{\text{imp}}=\sum_{\mu\sigma} \hat V_{\text{imp}}^{\mu\mu} c_{\mathbf R^* \mu  \sigma}^\dagger  c_{\mathbf R^* \mu  \sigma},
 \label{eq_H_imp}
\end{equation}
where the $V_{\text{imp}}^{\mu\mu}$ are on-site potentials as calculated from first principles,
we first construct
the lattice Green's functions in the presence of the impurity via the T-matrix approach
\begin{equation}
 \underline{\hat G}_{\mathbf{R}, \mathbf{R'}}(\omega)=\underline{\hat G}_{\mathbf{R}-\mathbf{R'}}^0(\omega)+\underline{\hat G}_{\mathbf{R}}^0(\omega)\underline{\hat T}(\omega)\underline{\hat G}_{-\mathbf{R'}}^0(\omega)\,.
 \label{eq_lattice_GF_T}
\end{equation}
The T-matrix is obtained from
\begin{equation}
 \underline{\hat T}(\omega)=[1-\underline{\hat V}_{\text{imp}}\underline{\hat G}(\omega)]^{-1} \underline{\hat V}_{\text{imp}},
 \label{eq_T_matrix}
\end{equation}
with the local Green's function $\underline{\hat G}(\omega)=\sum_{\mathbf{k}}\underline{\hat G}(\mathbf{k},\omega)$.
For the TM impurities, the corrections to the nearest-neighbor hopping and potentials
are generally small, e.g., one order of magnitude smaller such that we do not include
them here.
In contrast to fully self-consistent BdG calculations in real space (Refs. \onlinecite{Uranga16,Choubey14, gastiasoro_impurity_2013}),
the local suppression of the superconducting gap is not captured by the calculation, but
much larger spectral resolution can be achieved with reasonable computational effort.

The local lattice DOS in the presence of the impurity is
\begin{equation}
 \rho(\mathbf{R},\omega)=-\frac 1\pi \mathop{\text{Im}}\mathop {\text{Tr'}}\underline{\hat G}_{\mathbf{R}, \mathbf{R}}(\omega)\,,
 \label{eq_lattice_ldos}
\end{equation}
where $\mathop {\text{Tr'}}$ is the orbital trace over the normal part of the Nambu Green's function.

\begin{figure}[tb]
 \includegraphics[width=\linewidth]{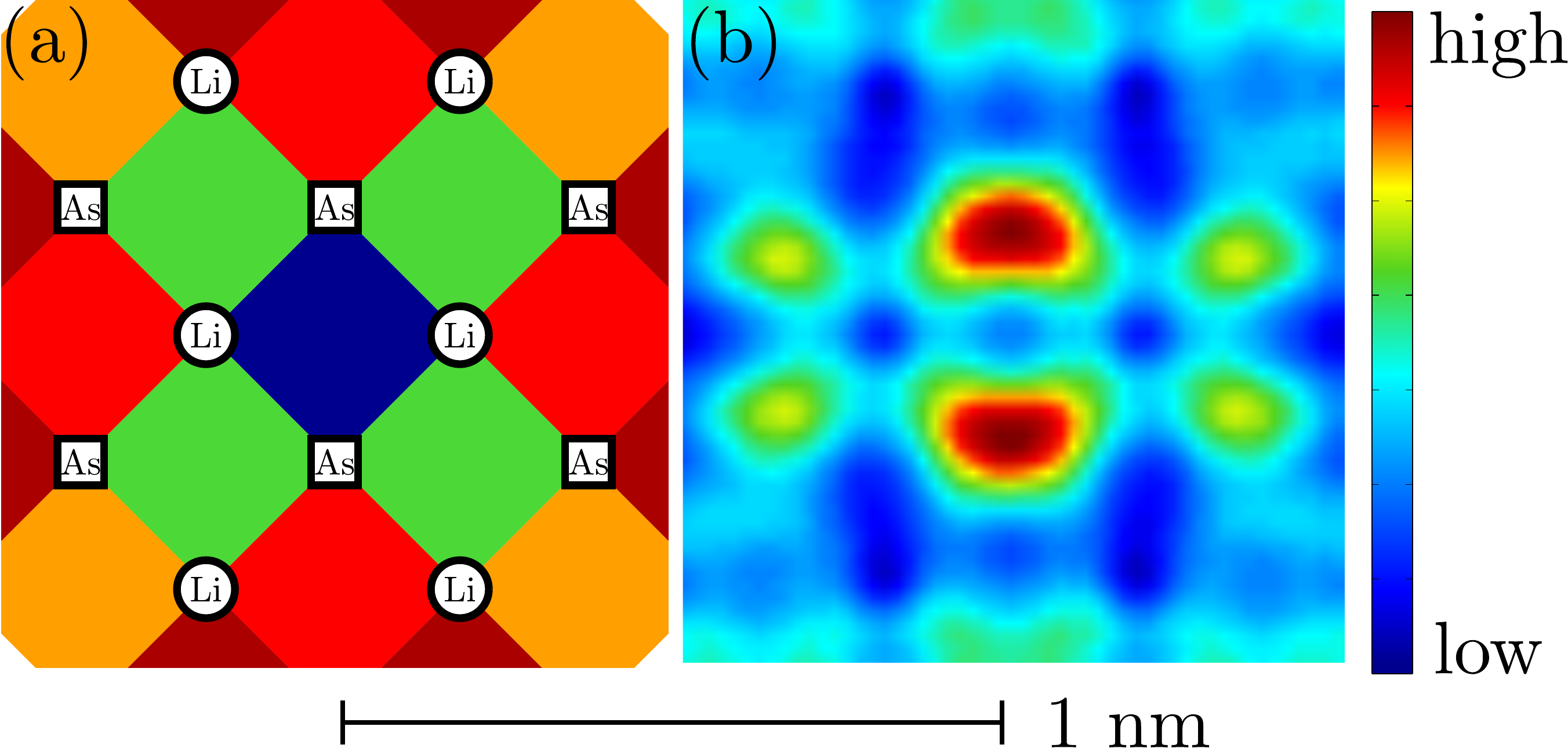}
\caption{(a) Lattice LDOS at $\omega/Z=-30\,\text{meV}$ around a Ni impurity showing a C$_4$ symmetric impurity state and the positions of the surface As and Li atoms in the field of view. (b) Corresponding cLDOS with the same field of view of $1\times 1 \,\text{nm}^2$ at $z=340\text{pm}$ reflecting the symmetry to be seen in conductance maps and topography close to a Fe centered impurity.}
\label{fig_maps_Ni}
\end{figure}

\begin{figure}[tb]
 \includegraphics[width=\linewidth]{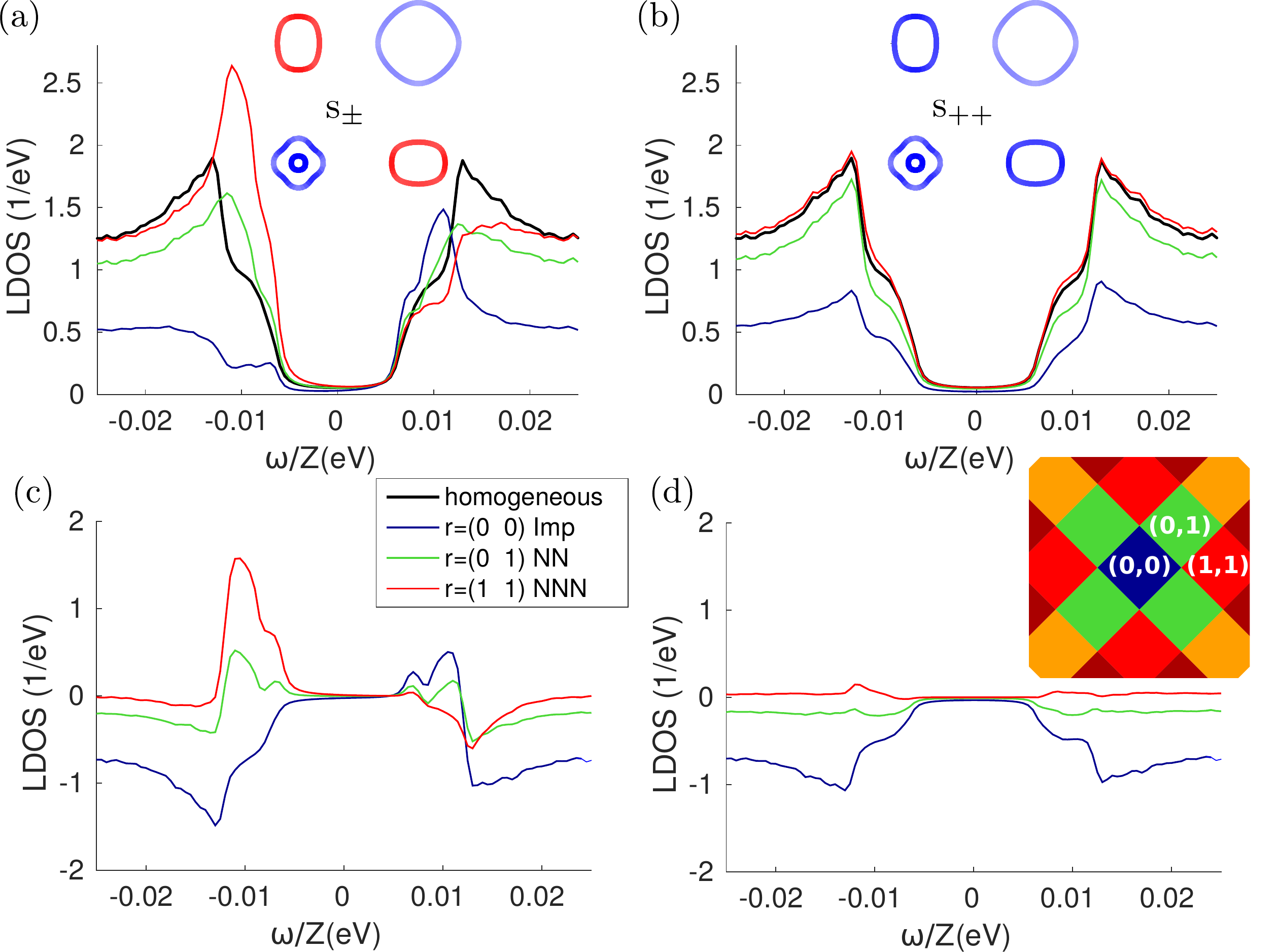}
\caption{Lattice LDOS spectra close to a Ni impurity (potential renormalized down by a factor of $2.5$) calculated in the lattice representation where $r=(x,y)$ refers to the distance in Fe bond lengths from the impurity. Panel (a) shows the result for the sign-changing gap function calculated from spin-fluctuation pairing theory, as plotted in the inset over the Fermi surface  using colors blue=+, red=-, with (c) the associated difference LDOS with spectrum far from the impurity subtracted.     Panel (b) is the corresponding LDOS result for an analogous  gap structure where the signs have been artificially removed (see Fig. \ref{fig_gap}(b)), showing no in-gap
bound states. The associated spectra are quite particle-hole symmetric, as seen in the LDOS difference spectrum in (d).
}
\label{fig_spectra}
\end{figure}

A plot of $\rho(\mathbf{R},\omega)$ at the resonance energy for a Ni impurity  is now shown in Fig. \ref{fig_maps_Ni}(a).  Because the underlying lattice Hamiltonian has $C_4$ symmetry, the pointlike impurity state inherits this symmetry.  On the other hand, when the cLDOS $\rho({\mathbf{r}},\omega)$ is calculated at the same energy, the characteristic ``geometric dimer'' state with maximum intensity on the NN As sites is immediately recovered\cite{Choubey14}, as evident from Fig. \ref{fig_maps_Ni}(b).  Note that this conductance map of the impurity resonance is obtained in an $s_\pm$ state calculated within the spin-fluctuation framework described in the previous section, as indicated in the inset of Fig.  \ref{fig_spectra} (a), where the corresponding spectra at various near neighbor sites of the Ni impurity are also shown.

Within this framework, it is also possible to show the effect of a sign-changing order parameter: the $s_\pm$ state exhibits in-gap bound states.  These are easiest to identify via difference spectra (impurity spectrum subtracted from the
homogeneous result) that are strongly asymmetric in energy, see Fig.\ref{fig_spectra}(c). A calculation where the
sign change of the order parameter has been artificially removed lacks the in-gap bound states and consequently shows
a nearly particle-hole symmetric difference spectrum as seen from Fig. \ref{fig_spectra}(d), i.e.,  allowing one to distinguish experimentally the two scenarios for
non-magnetic scatterers.


\begin{figure}[tb]
\includegraphics[width=\linewidth]{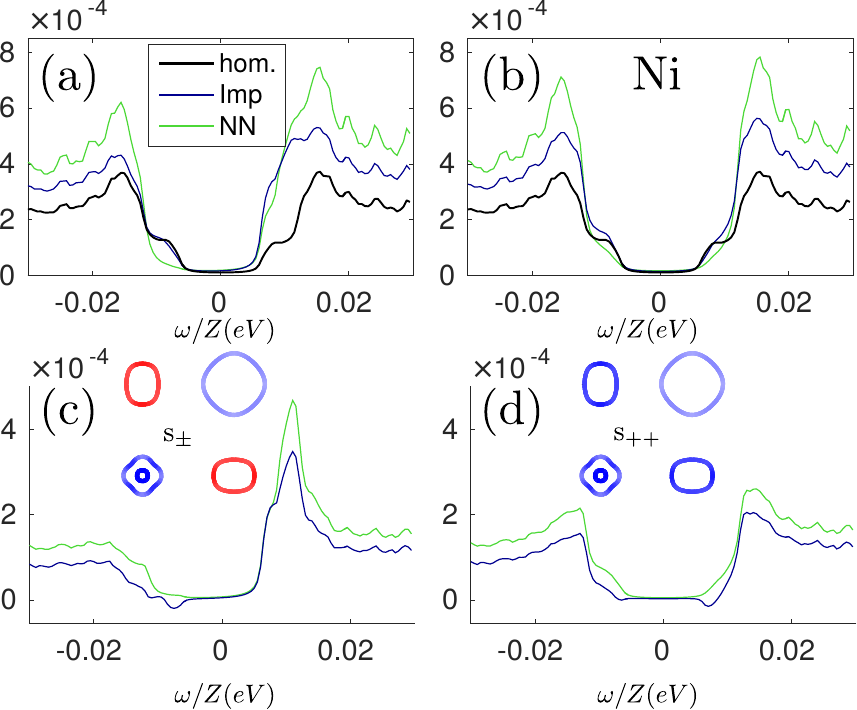}
\caption{Comparison of cLDOS results for a non-sign-changing order parameter (right) and a sign-changing order parameter (left): cLDOS close to a Ni impurity (top row) and the relative continuum LDOS (bottom row).}
\label{fig_spectra_spm_spp}
\end{figure}

There are other important differences between the lattice and continuum representations.  We stress that the impurity spectra themselves can be modified by
taking into account the tunneling process more precisely \cite{Tersoff1985,Kreisel15,Demler16}.
For example, looking at spectra obtained from Eq.~(\ref{eq_lattice_ldos}) for various positions around
a Ni impurity in an $s_\pm$ state, reveals that the effect of the impurity is much stronger on the negative
bias than on positive bias\cite{gastiasoro_impurity_2013}, as it can be seen best in the difference spectra Fig. \ref{fig_spectra}(c).
This feature changes when the continuum Green's functions are used to calculate the tunneling current. In Fig. \ref{fig_spectra_spm_spp}, we present
the spectra for the same impurity potential of a Ni impurity, but now calculated using Eq. (\ref{eq_basis}), where the changes at negative energies
are very small, but at positive energies a peak in the difference spectra (b) is obvious. On the other hand, the conclusions from
the calculation for the non-sign-changing order parameter do not change, e.g., the spectra do not show any indication
of in-gap bound states and consequently the difference spectra, Fig. \ref{fig_spectra_spm_spp} (d), are particle-hole
symmetric as expected. Experimentally, weak bound states at positive bias have been observed for such Ni impurities \cite{Chi2016}
giving evidence for a sign changing order parameter.

\begin{figure}[tb]
\includegraphics[width=\linewidth]{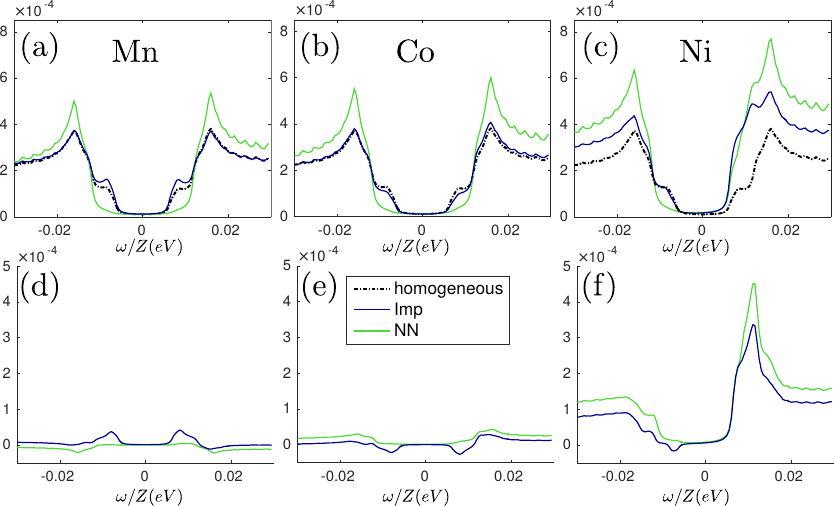}
\caption{Continuum LDOS close to a Mn impurity (a), Co impurity (b), and Ni impurity (c) and the corresponding relative spectra (d-f) calculated for the sign-changing order parameter. All impurity potentials reduced by 2.5 relative to {\it ab initio} calculations.}
\label{fig_spectra_cont}
\end{figure}

To make further connection to experimental investigations using known impurities in LiFeAs, we show a set of
calculated impurity spectra for the different substituents assuming a $s_\pm$ order parameter, see Fig. \ref{fig_spectra_cont}.
In all cases, the impurity potentials have been taken from our {\it ab initio} investigation, but divided by the same factor of 2.5\footnote{Note added after acceptance: In this work, the electronic structure was assumed to be
completely coherent, thus the requirement of reducing the impurity potentials could have the following explanation: One consequence of correlations in the electronic structure is a reduced quasiparticle weight $Z<1$ such that the Green's function in Eqs. (14) and (15) acquire an additional prefactor $G(\k,\omega) \rightarrow ZG(\k,\omega)$ within the Fermi liquid description\cite{Kreisel2016}. Not considering differences in orbitals, the inclusion of $Z=1/2$ in the Green's function yields the same result for the product of T-matrix and Green’s function as reducing the impurity potentials by an overall factor of $2$, see Eq. (15).}, see Appendix \ref{appendix_dft}.
It can be seen that the Mn impurity has opposite effect than the Co impurity which is due to the different sign of the impurity potential (d,e),
while the Ni impurity is stronger and shows an enhancement of the inner gap coherence peak, an effect that is consistent with
experimental evidence and can be understood in terms of the multiorbital nature of the electronic structure \cite{Chi2016}.

The method outlined so far was only considering non-magnetic impurity potentials, but it is straightforward to include
a magnetic scatterer in the classical approximation by making the potential spin-dependent, i.e., replacing $V_{\text{imp}}^{\mu\mu}\rightarrow V_{\sigma,\text{imp}}^{\mu\mu}$
in Eq. (\ref{eq_H_imp}) and doing the calculation for each spin species separately, eventually summing up the two contributions to simulate
a non-spin polarized experimental setup. At this point, we do not intend to present any results for a magnetic scatterer, because in a phenomenological
approach, more free parameters appear and make the discussion more complicated. In addition we expect the non-magnetic potentials to dominate over potential magnetic contributions, especially for Co and Ni ions.
From the experimental point of view, there are no indications that Co and Ni carry a significant magnetic moment. As for Mn impurities, on the other hand, they are expected to be magnetic\cite{Texier12,LeBoeuf14,Gastiasoro14} but only weakly coupled to the itinerant electron spins and hence one expects their magnetic potential to not strongly modify the LDOS\cite{Rosa14}.

\begin{figure}[tb]
\includegraphics[width=\linewidth]{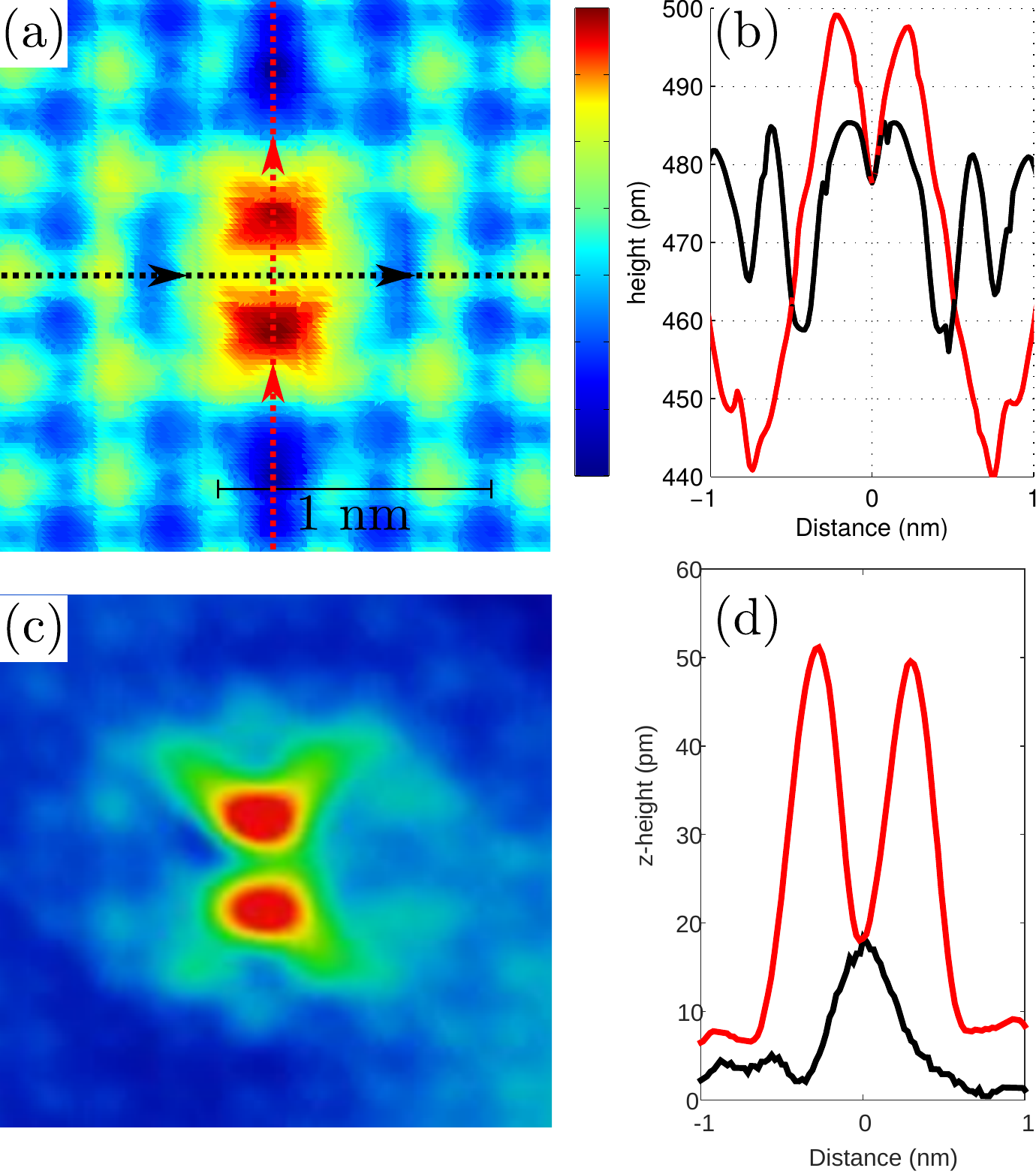}
\caption{Simulated topograph (a) for a Ni impurity at a bias of $30 \,\text{mV}$ calculated using a shifted Wannier function at the impurity position to approximate the distortion close to the impurity (see text)
and corresponding cut along the dashed lines in vertical and horizontal direction (b). (c) Experimental STM topograph of a single Ni impurity at $V_s=1 0\mathrm{mV}$, $I=500\mathrm{pA}$. (d) Line cuts along and normal to the Ni impurity shown in (c).}
\label{fig_topography_Ni_large_bias}
\end{figure}

Finally, we discuss consequences of our basis transformation for the spatial
modulations of the tunneling current, spectra and topographs close to impurities, e.g., using 
Eq.~(\ref{eq_lattice_GF_T}) to calculate the continuum Green's function in Eq.~(\ref{eq_basis}) to obtain the cLDOS following Eq.~(\ref{eq_cLDOS}).
Theoretical results for simulations of topographs close to a Ni impurity are given in Fig.~\ref{fig_topography_Ni_large_bias}.  The topographs $z(x,y)$ are calculated from Eq.~(\ref{eq:topograph}), and give our estimate for absolute height variations given a tip height and setpoint bias, as shown. Here local distortions of the atomic positions due to the impurity
have been taken into account by simply moving the Wannier function (from the homogeneous calculation) $\Delta z=20\,\text{pm}$ upwards. Without this adjustment by hand, we find that the topographs for tip heights several \AA~ above the plane do not provide sufficient contrast around the impurity site relative to the surrounding homogeneous surface. We believe that this  feature of the current calculation can be improved by the use of inhomogeneous Wannier functions, but this is beyond the scope of the current project.  

While the  height modulations through the impurity site shown in Fig. \ref{fig_topography_Ni_large_bias}  are of the order of experimental values, they are still larger by a factor of 2 compared to experiment in the case of the near-impurity modulations, and an order of magnitude larger than experiment far from the impurity site.  This results from a tip height that is probably too low compared to the real experiment; we are prevented from going higher by the noise in the exponential tails of the Wannier functions whose calculation was  described in Ref. \onlinecite{Berlijn11}.
As discussed earlier, the position of the maxima  in a topograph for the homogeneous case can depend on the bias voltage and current, but
the orientation of the impurity dimer appears always towards the NN As positions in our calculation.  In the case of the Ni dimer, the maxima are located close to the positions of the NN ``up"  As atoms, as in experiment.

\section{Discussion}
\label{sec:discussion}
We have simulated STM spectra and the corresponding topographs, accounting for
wave function information in real space together with the electronic structure from first-principles
calculations. Of course, it is well known that while {\it ab initio} approaches capture qualitative features of the electronic structure of Fe-SC, they cannot accurately describe many important near-Fermi-level properties
of Fe-SC in general\cite{Biermann16} and
LiFeAs in particular\cite{Ferberetal12,Haule12}. These uncertainties represent some of the underlying reasons that  the pairing state in this material is still under debate. The results of our theoretical calculations deriving from DFT-based band structures agree
quite well in many qualitative respects with the experimental findings, but clearly do not represent a complete quantitative solution to the problem.  In this section, we list discrepancies and possible explanations
together with proposals to investigate them more deeply.

As mentioned earlier, the impurity potentials obtained from a first-principles calculation
are too strong in magnitude to yield the weak in-gap bound states at the lower gap edge observed experimentally.
Since the identities of the impurities in these studies are well known, and they are well isolated, a one-impurity problem for the given chemical substituent is appropriate.  It seems likely therefore that the {\it ab initio} method simply
overestimates the impurity potentials\cite{Gastiasoro_2016} {\it relative to} the actual electronic structure. Assuming a screening that is present for all TM impurities
in a similar way, the impurity potentials were multiplied with the same overall renormalization,
ultimately  giving  a reasonable agreement of the spectra.  Thus  the relative strengths
seem to be calculated correctly, together with the signs of the various potentials.

Remaining discrepancies may also be due to an inadequate treatment of electronic correlations in the LiFeAs system.  While ideal from the point of view of experimental conditions for STM, LiFeAs near-Fermi-level electronic structure shows significant deviations from DFT, and current methods such as LDA+DMFT are unable to properly account for it  \cite{Ferberetal12,Haule12,Borisenko_12}.  The consequences of this discrepancy for the superconducting gap in spin-fluctuation theory were discussed at some length in Ref. \onlinecite{Hirschfeld_CRP_16}.  For the moment, the current approach is the best we have, but a truly quantitative theory awaits a proper calculation of the electronic structure in this system. 

Finally, we note that even with the renormalizations to account for the proper electronic structure introduced above there are quantitative discrepancies in the magnitudes of the observed oscillations of the tip near impurities.  We believe that accounting for local structural relaxations around the impurities and the use of an inhomogeneous Wannier basis can improve the description quantitatively. To some extent similar issues arise in our treatment of the homogeneous system, indicating that a more accurate treatment of the far field of the Wannier functions is also necessary.

\section{Conclusions}

We have presented a combined theoretical and experimental analysis of the superconducting state of LiFeAs and the modification of the LDOS due to defects. The theory is able to explain many, if not all of the features found in experiments.  The theory is based on a spin-fluctuation pairing prediction for the superconducting state using first-principles calculations for the electronic structure as input.  The resulting pairing gap has $s_\pm$ character, with the small gap located on the $d_{xy}$ band and the large gap on the $d_{xz/yz}$ bands, in good agreement with experiments.  Impurity potentials were also calculated within  first-principles approach.  The continuum local density of states for a single defect embedded at the LiFeAs surface was calculated using  recently developed Wannier-function based methods.

A surprising result of the theoretical study was that conductance maps and topographs of the  homogeneous system were found to be sensitive to tip height and setpoint bias, such that under different circumstances either the As or Li lattice could correspond to  intensity maxima in the image.   This preliminary study suggests caution in assigning site identification without a complete study varying experimental conditions.
Experimentally, impurity states are suitable to identify the correct positions of the Fe lattice and thus allow for the observation of a movement of the registered maxima in topographs depending on current and bias voltage. This effect is also predicted in our present framework and allows a more close comparison between theory and experiment to investigate more complicated impurity configurations or possibly unravel exotic electronic orders.
Comparison of topographic images as well as spectra of defects show that {\it ab initio} impurity potentials have to be renormalized by a factor of 2.5 to yield agreement with experiments.

Understanding the existence of the dimer peaks at positive bias requires the use of the Wannier-based analysis and is not consistent with conventional calculations of the Fe lattice LDOS. The calculations may only properly account for these when introducing relaxation effects around the impurity.
The good qualitative agreement of theory and experiment reported here bodes well for a future true quantitative analysis of inhomogeneous superconductivity and STS spectra in these systems.
We discussed improvements in  theoretical methods that are needed to reach this goal, as well as further experimental tests that may be important.

\begin{acknowledgments} The authors acknowledge useful discussions with C. Hess, Y. Wang, and D. Guterding. A.K. and B.M.A. acknowledge support from a Lundbeckfond fellowship (Grant No. A9318). S.C., D.B. and P.W. acknowledge funding from the MPG-UBC center. P.W. acknowledges financial support from EPSRC (EP/I031014/1). P.J.H. was supported by NSF-DMR-1407502.
A portion of this research was conducted at the Center for Nanophase Materials Sciences, which is a Department of Energy (DOE) Office of Science User Facility.
This manuscript has been authored by UT-Battelle, LLC under Contract No. DE-AC05-00OR22725 with the U.S. Department of Energy. 
W. K. acknowledges support from National Natural Science Foundation of China 11674220 and 11447601, and Ministry of Science and Technology 2016YFA0300500 and 2016YFA0300501.
Underpinning data can be obtained at \url{http://dx.doi.org/10.17630/ced13c7f-c9b6-479c-9668-e3d6c86775bc}.
\end{acknowledgments}
\appendix
\section{Electronic structure}
\label{appendix_dft}
For the construction of our model using {\it ab initio} methods, two first-principles calculations have been done:
In order to obtain a reasonable description of the electronic structure (bands) which eventually leads to
superconductivity, a calculation for a bulk system was done using experimental crystal structure of LiFeAs as reported in Ref.~\onlinecite{Tapp08} with
symmetry group $ P4/nmm$ and lattice constants $a=b=7.164\,\text{bohr}$, $c=12.026\,\text{bohr}$ as well as the fractional position $z= 0.2365$ of the As atoms.

\begin{figure}[tb]
 \includegraphics[width=\linewidth]{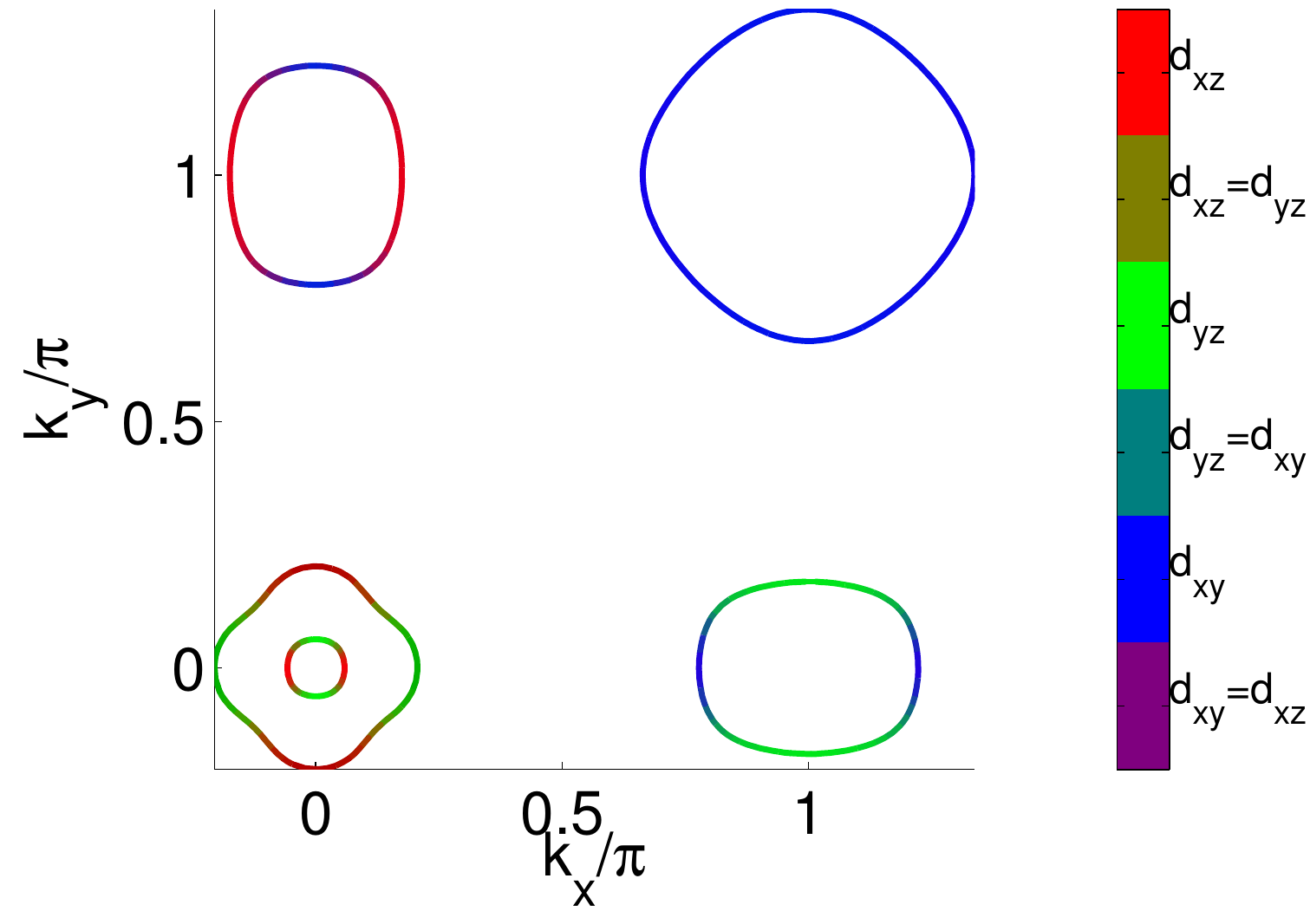}
\caption{Orbital character of the Fermi surface shown in $k$-space (set the lattice constants $a=c=1$) plotted in the unfolded (1-Fe) reciprocal unit cell, visualized with the summed-color method where the absolute value of the overlap is mapped to the RGB value of the color.}
\label{fig_fermi}
\end{figure}

After a Wannier projection, we map the 10 band model at $k_z=0$ onto a 5 band model via a standard gauge transformation \cite{Lee08,Eschrig09,Andersen11,Casula13,Fischer13,Cvetkovic13}, resulting in Eq. (\ref{eq_tb}). The mapping from 10 orbitals to a 5 orbital model is exact in this plane such that the Fermi surface of the 5 orbital model and its orbital character as shown in Fig.~\ref{fig_fermi} matches the corresponding Fermi surface of the 10 orbital model when plotted in the 2 Fe zone (not shown).
Next, a calculation of a monolayer LiFeAs is performed to obtain a set of Wannier functions describing the
electronic densities in the vacuum above the layer of Li and As atoms. For this purpose, the elementary cell
is extended in the $z$-direction by $20\,\text{bohr}$ to allow for a reasonably large vacuum between the two Li As layers.
As inferred from the partial density of states projected on the atoms, one sees that
the electronic states close to the Fermi level that are considered in our model
are dominantly of Fe $d$ character with small weight of As $p$ states. These contributions
express themselves in the lobes of the Wannier functions showing hybridization with As $p$ states.
On the other hand, Li states are negligible at low energies and thus do not hybridize
significantly with the Wannier functions, e.g., the numerical value of the Wannier function is
not enhanced close to the Li atoms unlike what happens close to the As atoms.

\section{Impurity potentials from {\it ab initio} approach}
The impurity potentials for the transition-metal ions were obtained from an {\it ab initio} supercell calculation as outlined in Ref. \onlinecite{Berlijn11}.
For completeness, we cite in Table \ref{table_Imp} the values of these potentials (on-site) orbital resolved. These have been multiplied with a factor of $0.4$ for the calculations
as presented in the main text.
\begin{table}[tb]
\centering
\begin{tabular}{c c c c c c}
 \hline
 Impurity & $d_{z^2}$ & $d_{x^2-y^2}$ & $d_{yz} =d_{yz}$  & $d_{xy}$ \\ [0.5ex]
 \hline\hline
 Mn          &0.315126  &   0.261420    &  0.271584 &  0.24206 \\
 Co           &   -0.371654 &  -0.30793    & -0.310513   &  -0.305179  \\
 Ni           & -1.56574 &   -1.86283      & -1.84138  &  -1.68037   \\
 \hline
\end{tabular}
\caption{Orbital resolved on-site impurity potentials (unrenormalized, in eV) of the transition-metal ions as obtained from first-principles calculations.}
\label{table_Imp}
\end{table}

\subsection{Spin-fluctuation theory}
\label{appendix_spin_fluc}
\begin{figure}[tb]
\includegraphics[width=1\columnwidth]{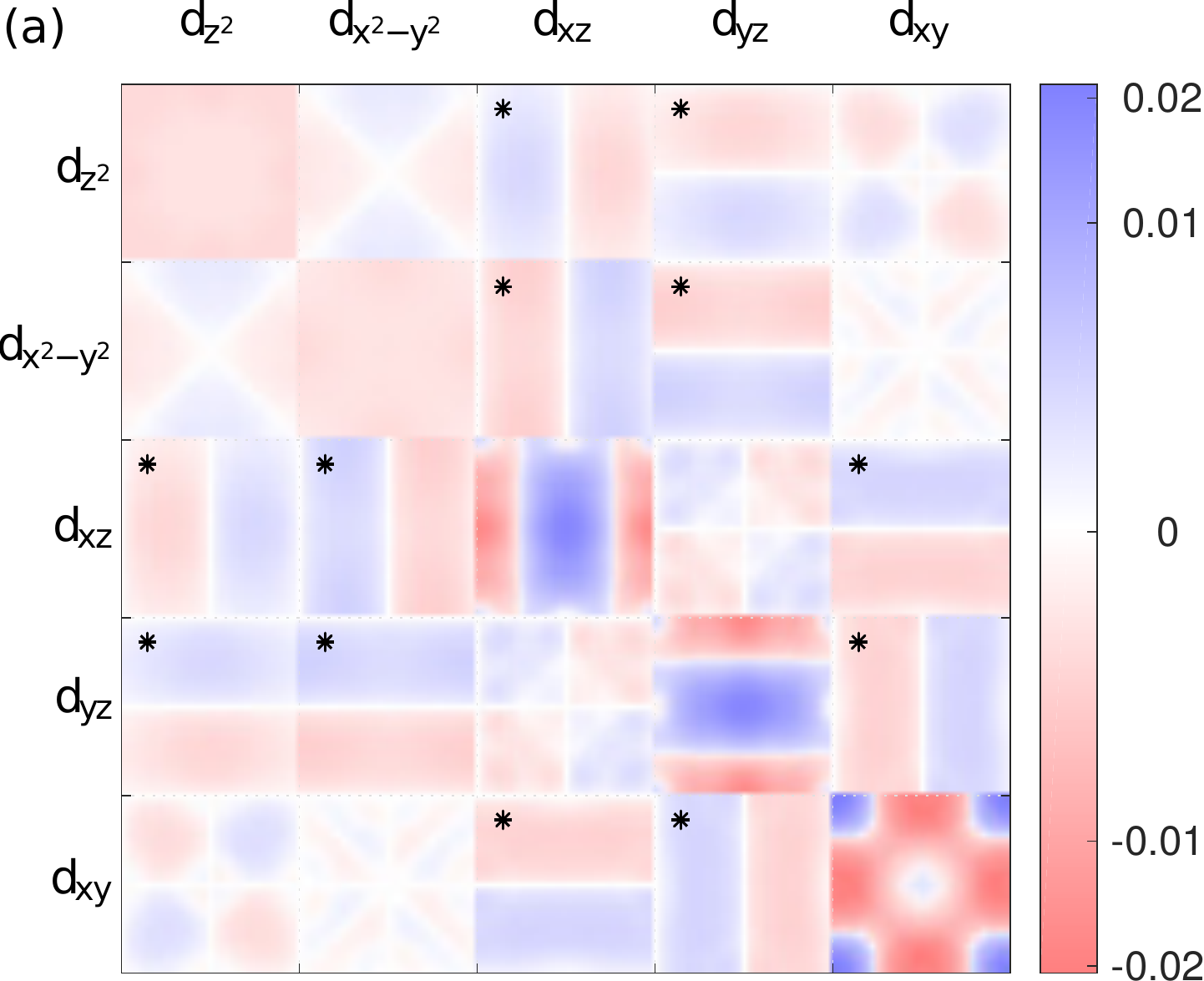}
\includegraphics[width=1\columnwidth]{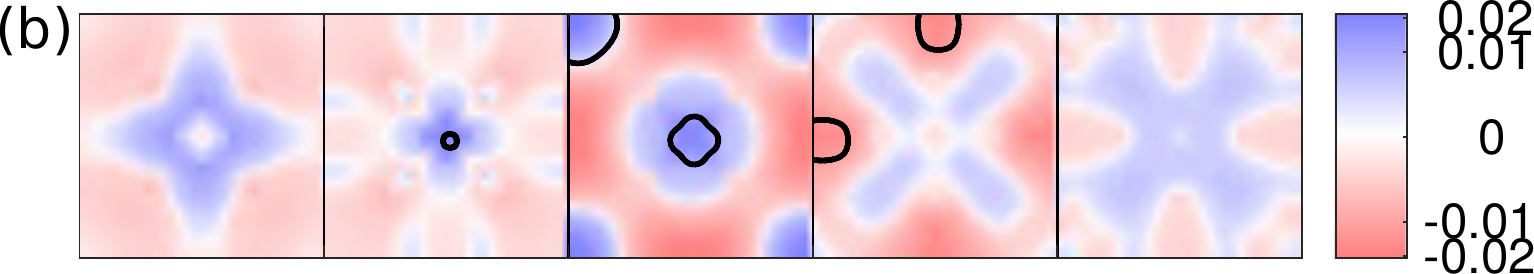}\newline
\caption{Converged superconducting gaps $\hat\Delta(\mathbf{\k})^{nm}$ (units in eV) from the BdG calculation at $T=1\text{meV}$ in orbital space (a) and transformed to band space (b). One $nm$ element in each square represents
the 1-Fe Brillouin zone with the $\Gamma$ point at the center. These gaps are calculated using the interaction parameters $U=0.9\,\text{eV}$ and $J=U/4$ and are used for all subsequent real-space calculations. The elements marked with a star $*$ are imaginary.
In the band space representation , the corresponding Fermi surfaces are marked in black in part of the Brillouin zone, compare Fig. \ref{fig_fermi}.}
\label{fig_delta}
\end{figure}
Next, we calculate the pairing interactions in momentum space following Ref. \onlinecite{a_kemper_10}, first adding the Hubbard Hund interaction to the Hamiltonian,
\begin{eqnarray}
	H =&{U}\sum_{\mathbf{R},\mu}n_{\mathbf{R}\mu\uparrow}n_{\mathbf{R}\mu\downarrow}+{U}'\sum_{\mathbf{R},\nu'<\mu}^\prime n_{\mathbf{R}\mu}n_{\mathbf{R}\nu}
	\nonumber\\
	 + & {J}\sum^\prime_{\mathbf{R},\nu<\mu}\sum_{\sigma,\sigma'}c_{\mathbf{R}\mu\sigma}^{\dagger}c_{\mathbf{R}\nu\sigma'}^{\dagger}c_{\mathbf{R}\mu\sigma'}c_{\mathbf{R}\nu\sigma}\\
	 + & {J}'\sum^\prime_{\mathbf{R},\nu\neq\mu}c_{\mathbf{R}\mu\uparrow}^{\dagger}c_{\mathbf{R}\mu\downarrow}^{\dagger}c_{\mathbf{R}\nu\downarrow}c_{\mathbf{R}\nu\uparrow}, \nonumber
\end{eqnarray}
where the interaction parameters ${U}$, ${U}'=U-2J$, ${J}$, ${J}'=J$ are expressed  in the notation of Kuroki \textit{et al.} \cite{Kuroki08}.
Assuming spin rotational invariance, we now calculate the pairing vertices in momentum space as
\renewcommand{\ell}{\mu}
\begin{align}
	&{\Gamma}_{\ell_1\ell_2\ell_3\ell_4} (\k-\k') = \left[\frac{3}{2} \bar U^s \chi_1^\text{RPA} (\k-\k') \bar U^s\nonumber \right.\,~~~~~~\,\\
	&\,~~~\left. +  \frac{1}{2} \bar U^s - \frac{1}{2}\bar U^c \chi_0^\text{RPA} (\k-\k') \bar U^c + \frac{1}{2} \bar U^c \right]_{\ell_1\ell_2\ell_3\ell_4}, \label{eq:fullGamma}
\end{align}
where the charge and spin susceptibilities are treated in the random phase approximation (RPA),
\begin{subequations}
\begin{align}
\chi_{1\,\ell_1\ell_2\ell_3\ell_4}^\text{RPA} (\q) &= \left\{ \chi^0 (\q) \left[1 -\bar U^s \chi^0 (\q) \right]^{-1} \right\}_{\ell_1\ell_2\ell_3\ell_4},\\
 \chi_{0\,\ell_1\ell_2\ell_3\ell_4}^\text{RPA} (\q) &= \left\{ \chi^0 (\q) \left[1 +\bar U^c \chi^0 (\q) \right]^{-1} \right\}_{\ell_1\ell_2\ell_3\ell_4},\label{eqn:RPA}
\end{align}
and the orbital-dependent susceptibilities have been calculated on a $k$ grid of $80\times 80$ together with a temperature broadening of $T_B=20\,\text{meV}$.
\end{subequations}
Definitions of the  bare orbital susceptibilities and interaction matrices $\bar U^s$ and $\bar U^c$ are given in Ref.~\onlinecite{a_kemper_10}.
The generalization of the symmetrized pairing interaction in the spin singlet channel for the multiorbital case in momentum space is then given by\cite{Roemer15}
\begin{equation}
 \Gamma_{\ell_1\ell_2\ell_3\ell_4}(\k,\k')=\frac 1 2 [{\Gamma}_{\ell_1\ell_2\ell_3\ell_4} (\k-\k')+{\Gamma}_{\ell_1\ell_2\ell_3\ell_4} (\k+\k')]\,.
\end{equation}

The result of Eq. (\ref{eq_gap_scf}) that has to be obtained self-consistently, is shown in Fig. \ref{fig_delta} where each orbital component of the order parameter is plotted in
the Brillouin zone. At first sight, one sees that the diagonal components (especially of the orbitals which are present on the Fermi level, $d_{xz}, d_{yz}$ and $d_{xy}$) are
largest, a consequence of the pairing interaction, Eq. (\ref{eq:fullGamma}), which is strongest in the intraorbital channel. We note further that individual components
reflect the symmetry relations of the orbitals, e.g., the $d_{xy}$, $d_z^2$ and $d_{x^2-y^2}$ orbitals show a $C_4$ symmetric order parameter while the order parameter in the $d_{xz}$ and $d_{yz}$ orbital
only has $C_2$ symmetry. The small off-diagonal elements, which are emphasized by the non-linear color scale, show similar properties, reflecting the transformation characteristics
of the participating orbitals, some of them purely imaginary. A transformation of the Hamiltonian given in Eq. (\ref{eq_H_nambu}) to the band basis by help of a unitary transformation that
diagonalizes $\hat H(\mathbf{k})$ yields the gap structure in band space shown in Fig. \ref{fig_delta} (b).
This normal-state transformation does not diagonalize the full Nambu Hamiltonian, thus  the off-diagonal terms (not shown) in band space are not zero indicating that our model also captures interband pairing. 
We are not discussing consequences of this property, but just mention that the gap structure in band space is the well know sign-changing $s_\pm$ with the gap on the k-points of the Fermi surface
in the normal state as presented in Fig. \ref{fig_gap}.

\label{Bibliography}
\end{document}